\documentclass[aps,prd,amsmath
,eqsecnum            % Adds section numbers to equation numbers.
,preprintnumbers
,nofootinbib         % Stops RevTeX moving footnotes to the references/bibliography
,twocolumn         % for two columns per page, as in the journal.  Only do this when close to submission.
]{revtex4}
\usepackage[dvips]{graphicx}% For importing graphics: .ps, .jpg, .bmp, .pcx, & others but NOT .gif .
\usepackage{amsmath}
\usepackage{amsfonts}
\usepackage{amssymb}
\usepackage{longtable}   % For tables that are too long for 1 page.

\allowdisplaybreaks[4]     % allows page breaks in the middle of an equation group.

\newcommand{\df}{\ {\overset {\rm def} =}\ }
\newcommand{\dr}[2]{\frac {{\rm d} {#1}} {{\rm d} {#2}}}
\newcommand{\pdr}[2]{\frac {\partial {#1}} {\partial {#2}}}
\newcommand{\dril}[2]{{{\rm d} {#1}} / {{\rm d} {#2}}}
\newcommand{\pdril}[2]{{\partial {#1}} / {\partial {#2}}}

\begin{document}

\title{Repeatable light paths in the shearfree normal cosmological models}

\author{Andrzej Krasi\'nski}
\affiliation{N. Copernicus Astronomical Centre, Polish Academy of Sciences, \\
Bartycka 18, 00 716 Warszawa, Poland} \email{akr@camk.edu.pl}

\date { }

\begin{abstract}
Conditions for the existence of repeatable light paths (RLPs) in the shearfree
normal cosmological models are investigated. It is found that in the conformally
nonflat models the only RLPs are radial null geodesics (in the spherical case)
and their analogues in the plane- and hyperbolically symmetric cases. In the
conformally flat Stephani models, there exist special spherically-, plane- and
hyperbolically symmetric subcases, in which all null geodesics are RLPs. They
are slightly more general than the Friedmann -- Lema\^{\i}tre -- Robertson --
Walker models of the corresponding symmetries: their curvature index function
$k(t)$ and the scale factor $R(t)$ are expressed through a single function of
time. In addition to that, there exist special cases of the Stephani solution in
which some of the null geodesics are RLPs. All these special cases are
identified.
\end{abstract}

\maketitle

\section{The motivation}

In a recent paper \cite{KrBo2011} it was found that in a general Szekeres model
\cite{Szek1975a, Szek1975b} of the $\beta' \neq 0$ class there are no repeatable
light paths (RLPs). This means, given a fixed light source S and a fixed
observer O, two light rays emitted from S in such a direction that they hit O,
but at different time instants, intersect different sequences  of matter world
lines on the way. The observer sees then the light source slowly drifting
through the sky. Depending on the mass distribution along the ray, and on the
positions of O and S with respect to each other, the average rate of change of
the direction toward S measured by O would be between $10^{-8}$ and $10^{-7}$
arc sec per year. With the current precision of direction determination equal to
$10^{-6}$ arc sec \cite{KrBo2011}, in the most favourable configuration 10 years
of monitoring would be required to measure this effect. Since this drift is
strictly zero in the Friedmann -- Lema\^{\i}tre -- Robertson -- Walker (FLRW)
models, it can be a qualitative observational test of homogeneity of the
Universe.

In Ref. \cite{KrBo2011} it was found that the only subcase of the $\beta' \neq
0$ Szekeres family of models in which {\em all} null geodesics are RLPs are the
Friedmann models. The condition for all null geodesics to be RLPs was the
vanishing of shear in the flow of the cosmic medium, which reduces the Szekeres
model to the Friedmann limit. In the general Szekeres models, the matter source
moves with zero acceleration (i.e. along timelike geodesics) and zero rotation.
This gives rise to the question whether the non-RLP phenomenon is caused by
shear or by the model being non-Friedmannian.

This question is addressed in the present paper. The condition of existence of
RLPs is applied here to the shearfree normal (SFN) cosmological models
\cite{Barn1973,Kras1997} -- another class of generalisations of the FLRW models,
in which the matter source is a perfect fluid moving with zero rotation and zero
shear, but nonzero acceleration. In the most general conformally nonflat SFN
models, the only RLPs are radial null geodesics (in the spherical case) and
their analogues in the plane- and hyperbolically symmetric cases. In the
conformally flat Stephani models, there exist special spherically-, plane-,
hyperbolically- and axially symmetric subcases, in which all null geodesics are
RLPs. They are slightly more general than the FLRW models of the corresponding
symmetries: their curvature index function $k(t)$ and the scale factor $R(t)$
are expressed through a single function of time. There also exist special cases
of the Stephani model, in which some of the null geodesics are RLPs. For
example, in the general axially symmetric case the RLPs are those null geodesics
that intersect each space $t =$ constant on the axis of symmetry (in full
analogy with the Szekeres models \cite{KrBo2011}). All these special cases are
identified.

Thus, the key to the RLP property is not vanishing shear, but high symmetry.

\section{The shearfree normal (SFN) cosmological models}\label{SFNmodels}

\setcounter{equation}{0}

The SFN models are solutions of Einstein's equations with a perfect fluid source
that moves with zero rotation and zero shear, but nonzero expansion and
acceleration. The full collection of these models was first found by Barnes
\cite{Barn1973} in 1973, but special cases were known before (see Ref.
\cite{Kras1997} for a classification of special cases and the account of
historical order). This family of models consists of the conformally flat
Stephani Universe \cite{Step1967} found in 1967, and 3 subfamilies of Petrov
type D solutions, first found in full generality by Barnes \cite{Barn1973}. Each
of these type D subfamilies has a 3-dimensional symmetry group acting on
2-dimensional orbits; the symmetry is either spherical or plane or hyperbolic.
The spherically symmetric subfamily was first presented by Kustaanheimo and
Qvist in 1948 \cite{KuQv1948}, but one special case of it was derived by
McVittie already in 1933 \cite{McVi1933}. The other two Petrov type D
subfamilies first emerged in the paper by Barnes \cite{Barn1973}. This author
\cite{Kras1989} found a coordinate system that covers all 3 subfamilies, but for
the present paper it will be more convenient to consider them separately.

In the Petrov type D case, the metric in comoving coordinates is
\begin{equation}
{\rm d} s^2 = \left(\frac {F V,_t} V\right)^2 {\rm d} t^2 - \frac 1 {V^2}
\left({\rm d} x^2 + {\rm d} y^2 + {\rm d} z^2\right), \label{2.1}
\end{equation}
where $F(t)$ is an arbitrary function, related to the expansion scalar $\theta$
by $\theta = 3 / F$. The Einstein equations reduce to the single equation:
\begin{equation}
w,_{uu} /w^{2} = f(u), \label{2.2}
\end{equation}
where $f(u)$ is another arbitrary function, while the variable $u$ and the
function $w$ are related to the coordinates $x, y, z$, and to the function
$V(t,x,y,z)$ differently in each subfamily. We have
\begin{eqnarray}\label{2.3}
(u, w) &=& \left\{ \begin{array}{lll} (r^{2}, V)
   & \mbox{with spherical symmetry};\\
(z, V)  & \mbox{with plane symmetry};\\
(x/y, V/y)  & \mbox{with hyperbolic symmetry}.\end{array} \right. \nonumber \\
r^2 &\df& x^2 + y^2 + z^2.
\end{eqnarray}
The formulae for matter density and pressure are known for each case, but will
not be used in the present paper, so they are not quoted; see Ref.
\cite{Kras1997}.

The Weyl tensor is proportional to $f(u)$, and so with $f(u) \equiv 0$ the
models given by (\ref{2.2}) become conformally flat. Then they become subcases
of the Stephani solution given below, but are still more general than FLRW, see
Ref. \cite{Kras1997}.

The conformally flat Stephani solution \cite{Step1967, Kras1997} has the metric
given by (\ref{2.1}), the coordinates are still comoving, but the function $V(t,
x, y, z)$ is given explicitly by
\begin{eqnarray}
&& V = \frac 1 R \left\{1 + \frac 1 4 k(t) \left[\left(x -
x_{0}(t)\right)^{2}\right.\right. \nonumber \\
&& \ \ \ \ \  + \left.\left.\left(y - y_{0}(t)\right)^{2} + \left(z -
z_{0}(t)\right)^{2}\right]\right\}, \label{2.4}
\end{eqnarray}
where $(R, k, x_0, y_0, z_0)$ are arbitrary functions of $t$. This is easily
seen to be a generalisation of the whole FLRW class, to which it reduces when
$(k, x_0, y_0, z_0)$ are all constant. The constants $(x_0, y_0, z_0)$ can then
be set to zero by a coordinate transformation, the constant $k$ is the FLRW
curvature index, and $R(t)$ is the FLRW scale factor. In general, the solution
(\ref{2.4}) has no symmetry.

As with the type D models, the formulae for mass density and pressure are known,
but will not be used here, see \cite{Step1967, Kras1997}. The mass density
depends only on $t$.

The parametrisation used in (\ref{2.4}) follows the original source
\cite{Step1967} and is designed so that the most popular representation of the
FLRW models is easily obtained from it as a spatially homogeneous limit. It
suggests that $V$ is either quadratic in $(x, y, z)$ or does not depend on them.
In fact, the metric (\ref{2.1}) is still conformally flat with $V = A_0 + A_q (x
^ 2 + y ^ 2 + z ^ 2) + A_1 x + A_2 y + A_3 z$, where the $A$'s are arbitrary
functions of $t$. However, even when $A_q = 0$, the quadratic terms can be
restored with use of the Haantjes transformation, see Refs. \cite{Kras1989,
PlKr2006} for the definition and examples of transformations.

As already mentioned, all these solutions have zero shear and zero rotation. The
quantity that makes them more general than FLRW is acceleration, which is in
every case proportional to the spatial gradient of the $g_{00}$ component in the
metric (\ref{2.1}). Thus, the invariant condition for each of the metrics
discussed above to reduce to an FLRW limit is
\begin{equation}\label{2.5}
\pdr {} {x^i} \left(\frac {V,_t} V\right) = 0, \quad i = 1, 2, 3.
\end{equation}

Note that the metrics (\ref{2.1}) do not allow a static limit -- $V,_t$ must be
nonzero. Static solutions that are shearfree and rotation-free obviously exist,
but they form a separate branch in the family of solutions of Einstein's
equations and cannot be recovered from (\ref{2.1}).

There exists a very large body of literature on the SFN models, their various
subcases and generalisations \cite{Kras1997}, but it is almost exclusively
devoted to finding examples of solutions of (\ref{2.2}) and of its charged
generalisation, $w,_{uu} = f(u) w^2 + g(u) w^3$. A general solution is not
known.

\part{\Large The spherically symmetric models}

\section{The geodesic equations}\label{sphergeo}

\setcounter{equation}{0}

For the spherically symmetric SFN models we first transform the spatial
coordinates of (\ref{2.1}) to the standard spherical coordinates. Then the
geodesic equations in the affine parametrisation are:
\begin{eqnarray}
&& \dr {^2t} {s^2} + \left(\frac {F,_t} F - \frac {V,_t} V + \frac {V,_{t
t}} {V,_t}\right) \left(\dr t s\right)^2 \nonumber \\
&&\ \ \ - \frac 1 {F^2 V V,_t} \left[\left(\dr r s\right)^2 + u \left(\dr
\vartheta
s\right)^2 + u \sin^2 \vartheta \left(\dr \varphi s\right)^2\right] \nonumber \\
&&\ \ \ + 2 \left(\frac {V,_{t u}} {V,_t} - \frac {V,_u} V\right) \dr t s \dr u
s = 0, \label{3.1} \\
&& \dr {^2r} {s^2} + 2 r F^2 V,_t \left(V,_{t u} - \frac {V,_t V,_u} V\right)
\left(\dr t s\right)^2 \nonumber \\
&&\ \ \ - 2 \frac {V,_t} V \dr t s \dr r s - r \left[\left(\dr \vartheta
s\right)^2 + \sin^2 \vartheta \left(\dr \varphi s\right)^2\right] \nonumber \\
&&\ \ \ + 2 r \frac {V,_u} V \left[- \left(\dr r s\right)^2 + u \left(\dr
\vartheta s\right)^2 + u \sin^2 \vartheta \left(\dr \varphi s\right)^2\right]
\nonumber \\
&&\ \ \ = 0, \label{3.2} \\
&& \dr {^2 \vartheta} {s^2} - 2 \frac {V,_t} V \dr t s \dr \vartheta s +
\left(\frac 1 u - 2 \frac {V,_u} V\right) \dr u s \dr \vartheta s \nonumber \\
&&\ \ \ - \cos \vartheta \sin \vartheta \left(\dr \varphi s\right)^2 = 0,
\label{3.3} \\
&& \dr {^2 \varphi} {s^2} - 2 \frac {V,_t} V \dr t s \dr \varphi s +
\left(\frac 1 u - 2 \frac {V,_u} V\right) \dr u s \dr \varphi s \nonumber \\
&&\ \ \ + 2 \cot \vartheta \dr \vartheta s \dr \varphi s = 0. \label{3.4}
\end{eqnarray}

Equations (\ref{3.1}) -- (\ref{3.2}) can be simplified when we use the null
condition:
\begin{equation}\label{3.5}
\left(\dr r s\right)^2 + u \left(\dr \vartheta s\right)^2 + u \sin^2 \vartheta
\left(\dr \varphi s\right)^2 = \left(F V,_t\right)^2 \left(\dr t s\right)^2.
\end{equation}
What remains of them is:
\begin{eqnarray}\label{3.6}
\dr {^2t} {s^2} &+& \left(\frac {F,_t} F - 2 \frac {V,_t} V + \frac {V,_{t
t}} {V,_t}\right) \left(\dr t s\right)^2 \nonumber \\
&+& 2 \left(\frac {V,_{t u}} {V,_t} - \frac {V,_u} V\right) \dr t s \dr u s = 0,
\\
\dr {^2r} {s^2} &+& \frac {F^2 V,_t} r \left(2 u V,_{t u} - V,_t\right)
\left(\dr t s\right)^2 - 2 \frac {V,_t} V \dr t s \dr r s  \nonumber \\
&+& \frac 1 r \left(1 - 4u \frac {V,_u} V\right) \left(\dr r s\right)^2 = 0.
\label{3.7}
\end{eqnarray}

{}From (\ref{3.3}) -- (\ref{3.4}) it follows that the null geodesics are plane,
i.e. that the coordinates can be adapted to each {\em single} geodesic so that
it has $\vartheta = \pi/2$ all along. However, in the following we will consider
a bundle of geodesics emanating from a common source, and such an adaptation of
coordinates would not be useful for that.

In this and the next two sections, the same method as in Ref. \cite{KrBo2011}
will be used, so, for readers' convenience, the relevant short excerpts from
there are copied here with suitable modifications.

For further calculations it is more convenient to use the coordinate $r$ as a
parameter, which will be non-affine. This is allowed, but with some caution. It
is easily seen from (\ref{3.7}) that a curve obeying it, on which $\dril r s =
0$ over some open range of $s$ has $\dril t s = 0$ in that range, and so is
spacelike (the second possibility, $2 u V,_{t u} - V,_t = 0$, could be
compatible with (\ref{2.2}) only in the prohibited case $V,_t = 0$). However,
(\ref{3.6}) -- (\ref{3.7}) do not guarantee that $\dril r s \neq 0$ at all
points; isolated points along a null geodesic, at which $\dril r s =0$ can
exist. Thus, $r$ can be used as a parameter on null geodesics only on such
segments where $\dril s r
> 0$ or $\dril s r < 0$ throughout.

We have, for any coordinate:
\begin{equation}\label{3.8}
\dr {^2 x^\alpha} {s^2} = \left(\dr r s\right)^2 \dr {^2 x^\alpha} {r^2} + \dr
{^2 r} {s^2} \dr {x^\alpha} r.
\end{equation}
Then, from (\ref{3.7}) we have:
\begin{eqnarray}\label{3.9}
&& \frac {{\rm d}^2 r} {{\rm d} s^2} = \left(\dr r s\right)^2 \left[- \frac {F^2
V,_t} r \left(2 u V,_{t u} - V,_t\right) \left(\dr t r\right)^2\right.
\nonumber \\
&& + \left.2 \frac {V,_t} V \dr t r - \frac 1 r \left(1 - 4u \frac {V,_u}
V\right)\right].
\end{eqnarray}
Consequently, (\ref{3.6}) and (\ref{3.3}) -- (\ref{3.4}) become, using
(\ref{3.9}):
\begin{eqnarray}\label{3.10}
\dr {^2 t} {r^2} &+& \left(4 r \frac {V,_{t u}} {V,_t} - \frac 1 r\right) \dr t
r + \left(\frac {F,_t} F + \frac {V,_{t t}} {V,_t}\right) \left(\dr t r\right)^2
\nonumber \\
&-& \frac {F^2 V,_t} r \left(2 u V,_{t u} - V,_t\right) \left(\dr t r\right)^3 =
0, \\
\dr {^2 \vartheta} {r^2} &+& \dr \vartheta r \left[- \frac {F^2 V,_t} r \left(2
u V,_{t u} - V,_t\right) \left(\dr t r\right)^2 + \frac 1 r\right] \nonumber
\label{3.11} \\
&-& \cos \vartheta \sin \vartheta \left(\dr \varphi r\right)^2 = 0, \\
\dr {^2 \varphi} {r^2} &+& \dr \varphi r \left[- \frac {F^2 V,_t} r \left(2 u
V,_{t u} - V,_t\right) \left(\dr t r\right)^2 + \frac 1 r\right] \nonumber
\label{3.12} \\
&+& 2 \cot \vartheta \dr \vartheta r \dr \varphi r = 0.
\end{eqnarray}

\section{The redshift equations}\label{spheredshift}

\setcounter{equation}{0}

Consider, in the metric (\ref{2.1}), two light signals, the second one following
the first after a short time-interval $\tau$, both emitted by the same source
and arriving at the same observer. The equation of the trajectory of the first
signal is
\begin{equation}\label{4.1}
(t, \vartheta, \varphi) = (T(r), \Theta(r), \Phi(r)),
\end{equation}
the corresponding equation for the second signal is
\begin{equation}\label{4.2}
(t, \vartheta, \varphi) = (T(r) + \tau(r), \Theta(r) + \zeta(r), \Phi(r) +
\psi(r)).
\end{equation}
Thus, we have to allow that while the first ray intersects the hypersurface of a
given constant value of the $r$-coordinate at the point $(t, \vartheta, \varphi)
= (T, \Theta, \Phi)$, the second ray intersects the same hypersurface at the
point $(t, \vartheta, \varphi) = (T + \tau, \Theta + \zeta, \Phi + \psi)$. In
general, it arrives at this hypersurface not only later, but also at a different
spatial location. Thus, those two rays will not intersect the same succession of
intermediate matter worldlines on the way. Note that, since the coordinates used
here are comoving, both the source of light and the observer keep their spatial
coordinates unchanged throughout history. Given this, and given that a pair of
rays emitted by the same source and received by the same observer is considered,
$(\zeta, \psi) = (0, 0)$ holds at the point of emission and at the point of
reception. However, the second ray is in general emitted in a different
direction than the first one, and is received from a different direction by the
observer. This means that in a general cosmological model the observed objects
should drift across the sky. (See a brief quantitative discussion of this effect
in the Lema\^{\i}tre -- Tolman model in Ref. \cite{KrBo2011}.) The directions of
the two rays will be determined by $(\dril \Theta r, \dril \Phi r)$ and $(\dril
\Theta r + \xi(r), \dril \Phi r + \eta(r))$, respectively, where $\xi = \dril
{\zeta} r$, $\eta = \dril {\psi} r$. It will be assumed here that $(\dril {\tau}
r, \zeta, \psi, \xi, \eta)$ are small of the same order as $\tau$, so all terms
nonlinear in any of them and terms involving their products will be neglected.

In writing out the equations of propagation of redshift, the symbol $\Delta$
will be used. It denotes the difference between the relevant expression taken at
$(t + \tau, r, \vartheta + \zeta, \varphi + \psi)$ and at $(t, r, \vartheta,
\varphi)$, {\em linearized in} $(\tau, \zeta, \psi)$; i.e. the difference
between the value of a given quantity along the second ray and along the first
ray, taken at a hypersurface of a given value of the parameter $r$ (and,
automatically, given $u$). For example $F(t + \tau) - F(t) = \Delta F + {\cal
O}(\tau^2)$, $V(t + \tau, u) - V(t, u) = \Delta V + {\cal O}(\tau^2)$, $\Delta
(\dril \vartheta r) = \xi$. This operation is a generalisation of the
calculation by which Bondi \cite{Bond1947} derived the redshift equation for
radial null geodesics in the Lema\^{\i}tre -- Tolman model, see an account of
that method in Ref. \cite{KrBo2011}.

Applying the $\Delta$ operation to (\ref{3.10}) -- (\ref{3.12}) we obtain the
following general equations of redshift propagation in the spherically symmetric
SFN models:\footnote{A quick way to calculate (\ref{4.3}) -- (\ref{4.6}) is to
take the differential of the corresponding quantity at constant $r$ and replace
$({\rm d} t, {\rm d} \vartheta, {\rm d} \varphi, {\rm d} (\dril \vartheta r),
{\rm d} (\dril \varphi r))$ by $(\tau, \zeta, \psi, \xi, \eta)$.}
\begin{eqnarray}
&& \dr {^2 \tau} {r^2} + 4 r \left(\frac {V,_{ttu}} {V,_t} - \frac {V,_{tt}
V,_{tu}} {{V,_t}^2}\right) \dr t r \tau \nonumber \\
&& + \left(\frac {F,_{tt}} F - \frac {{F,_t}^2} {F^2} + \frac {V,_{ttt}} {V,_t}
- \frac {{V,_{tt}}^2} {{V,_t}^2}\right) \left(\dr t r\right)^2 \tau \nonumber \\
&&\ \ \ - \frac F r \left[\left(2 F,_t V,_t + F V,_{tt}\right) \left(2u V,_{tu}
- V,_t\right)\right. \nonumber \\
&&\ \ \ \ \ \ + \left.F V,_t \left(2 u V,_{ttu} - V,_{tt}\right)\right]
\left(\dr t r\right)^3 \tau \nonumber \\
&&\ \ \ + \left(4 r \frac {V,_{t u}} {V,_t} - \frac 1 r\right) \dr \tau r + 2
\left(\frac {F,_t} F + \frac {V,_{tt}} {V,_t}\right) \dr t r \dr \tau r
\nonumber \\
&&\ \ \ - \frac {3 F^2 V,_t} r \left(2u V,_{tu} - V,_t\right) \left(\dr
t r\right)^2 \dr \tau r = 0, \label{4.3} \\
&& \dr \xi r + \left[- \frac {F^2 V,_t} r \left(2u V,_{tu} -
V,_t\right) \left(\dr t r\right)^2 + \frac 1 r\right] \xi \nonumber \\
&&\ \ \ - \frac F r \left[\left(2 F,_t V,_t + F V,_{tt}\right) \left(2u V,_{tu}
- V,_t\right)\right. \nonumber \\
&&\ \ \ \ \ \ + \left.F V,_t \left(2 u V,_{ttu} - V,_{tt}\right)\right]
\left(\dr t r\right)^2 \dr \vartheta r \tau \nonumber \\
&&\ \ \ - \frac {2 F^2 V,_t} r \left(2u V,_{tu} - V,_t\right) \dr t
r \dr \vartheta r \dr \tau r \nonumber \\
&&\ \ \ - \cos (2 \vartheta) \left(\dr \varphi r\right)^2 \zeta - \sin (2
\vartheta) \dr \varphi r \eta = 0, \label{4.4} \\
&& \dr \eta r + \left[- \frac {F^2 V,_t} r \left(2u V,_{tu} -
V,_t\right) \left(\dr t r\right)^2 + \frac 1 r\right] \eta \nonumber \\
&&\ \ \ - \frac F r \left[\left(2 F,_t V,_t + F V,_{tt}\right) \left(2u V,_{tu}
- V,_t\right)\right. \nonumber \\
&&\ \ \ \ \ \ + \left.F V,_t \left(2 u V,_{ttu} - V,_{tt}\right)\right]
\left(\dr t r\right)^2 \dr \varphi r \tau \nonumber \\
&&\ \ \ - \frac {2 F^2 V,_t} r \left(2u V,_{tu} - V,_t\right) \dr t r \dr
\varphi r \dr \tau r - \frac 2 {\sin^2 \vartheta} \dr \vartheta r \dr \varphi r
\zeta \nonumber \\
&&\ \ \ + 2 \cot \vartheta \dr \vartheta r \eta + 2 \cot \vartheta \dr \varphi r
\xi = 0. \label{4.5}
\end{eqnarray}
Along with these, we can use also the result of $\Delta$ acting on the null
condition (\ref{3.5}), transformed to the $r$-parametrisation, it is
\begin{eqnarray} \label{4.6}
&&\left(2 F F,_t {V,_t}^2 + 2 F^2 V,_t V,_{tt}\right) \left(\dr t r\right)^2
\tau + 2 F^2 {V,_t}^2 \dr t r \dr \tau r \nonumber \\
&& = 2 u \dr \vartheta r \xi + u \sin (2 \vartheta) \left(\dr \varphi r\right)^2
\zeta + 2 u \sin^2 \vartheta \dr \varphi r \eta.
\end{eqnarray}
If ${\cal T}$ is the period of the electromagnetic wave measured in the rest
frame of the source or of the observer, the redshift is given by
\begin{equation}\label{4.7}
\frac {{\cal T}(r_{\rm obs})} {{\cal T}(r_{\rm em})} = 1 + z(r_{\rm em}),
\end{equation}
where the labels ``obs'' and ``em'' refer to the events of observation and
emission of the ray, respectively \cite{KrBo2011, PlKr2006}. The period in the
rest frame of an object is measured in the units of proper time of that object.
In the metrics (\ref{2.1}) the differential of the proper time, ${\rm d} s$, is
related to the differential of the coordinate time, ${\rm d} t$, by ${\rm d} s =
\sqrt{g_{00}} {\rm d} t$, where $g_{00} = \left(FV,_t/V\right)^2$. Thus, taking
the $\tau(r)$ found from (\ref{4.3}) -- (\ref{4.6}) as corresponding to the
period ${\cal T}$, we calculate the redshift from
\begin{equation}\label{4.8}
1 + z(r_{\rm em}) = \frac {\left.\left(\sqrt{g_{00}} \tau\right)\right|_{t_{\rm
obs}, r_{\rm obs}}} {\left.\left(\sqrt{g_{00}} \tau\right)\right|_{t_{\rm em},
r_{\rm em}}}.
\end{equation}

Note that eq. (\ref{4.3}) is decoupled from the other two and determines the
redshift independently of $(\zeta, \psi, \xi, \eta)$. This means that even for a
non-radial ray the redshift changes with $r$ by the same law as for a radial
one. (Indeed, (\ref{3.6}) is the same for radial and non-radial rays, so
$\Delta$ acting on it gives the same result in both cases.) This is a
consequence of spherical symmetry {\em and zero shear} in the SFN models. Using
eqs. (4.1) -- (4.5) and (5.3) -- (5.12) of Ref. \cite{KrBo2011} one can verify
that in the Szekeres model the equation (5.12) that determines $\tau$ does
depend on $(\zeta, \psi, \xi, \eta)$ even in the spherically symmetric subcase,
where ${\cal E},_r = 0$ and shear is nonzero.

Since $\zeta = \psi = 0$ at the observer, these quantities are not in fact
observable. However, $\xi$ and $\eta$ are in general nonzero at the observer,
which implies the change of direction toward the source with time.

\section{Repeatable light paths (RLPs)}\label{spherlps}

\setcounter{equation}{0}

We say that the light paths are repeatable when the rays sent between a given
source and a given observer at different times always proceed through the same
sequence of intermediate particles of the cosmic medium. This means, when the
rays are registered at an $r$-hypersurface of coordinate radius $r$, they arrive
there at the same spatial location (only at different time instants), i.e.:
\begin{equation}\label{5.1}
\zeta = \psi = \xi = \eta = 0
\end{equation}
all along each ray. Substituting (\ref{5.1}) in (\ref{4.4}) -- (\ref{4.6}) one
obtains the conditions that have to be obeyed in order that RLPs exist. We first
define:
\begin{eqnarray} \label{5.2}
&& \chi(t, u) \df \dr t r \tau \left[\left(2 F,_t V,_t + F V,_{tt}\right)
\left(2u V,_{tu} - V,_t\right)\right. \nonumber \\
&&\ \ \ + \left.F V,_t \left(2 u V,_{ttu} - V,_{tt}\right)\right] + 2 F V,_t
\left(2u V,_{tu} - V,_t\right) \dr \tau r \nonumber \\
\end{eqnarray}
and then the conditions are
\begin{eqnarray}
&&\hspace{-2mm} - \frac F r \chi \dr t r \dr \vartheta r = 0, \label{5.3} \\
&&\hspace{-2mm} - \frac F r \chi \dr t r \dr \varphi r = 0, \label{5.4} \\
&&\hspace{-2mm}  2 F V,_t \left[\left(F,_t V,_t + F V,_{tt}\right) \dr t r \tau
+ F V,_t \dr \tau r\right] \dr t r = 0.\ \ \ \ \ \  \label{5.5}
\end{eqnarray}

Discarding in (\ref{5.5}) the impossible solutions $\dril t r = 0$ (this would
be a spacelike curve) and $F V,_t = 0$ (prohibited -- see (\ref{2.1})) we
obtain:
\begin{equation}\label{5.6}
\dr \tau r = - \left(\frac {F,_t} F + \frac {V,_{tt}} {V,_t}\right) \dr t r
\tau.
\end{equation}
Equations (\ref{5.3}) -- (\ref{5.4}) imply that either $\dril \vartheta r =
\dril \varphi r = 0$ or $\chi = 0$. The first case defines a radial null
geodesic. This means that radial null geodesics are RLPs, as expected in a
spherically symmetric model. Then (\ref{5.6}) just defines the redshift as a
function of $r$.

To check the other possibility we substitute (\ref{5.6}) in $\chi = 0$. The
result is, after discarding the factor $2 u \left(F^2/r\right) (\dril t r)
\tau$:
\begin{equation}\label{5.7}
V,_{tt} V,_{tu} - V,_t V,_{ttu} = 0.
\end{equation}
A general solution of this equation is
\begin{equation}\label{5.8}
V = \alpha(u) S(t) + \beta(u),
\end{equation}
where $\alpha$, $\beta$ and $S$ are arbitrary functions of their respective
arguments. This must be compatible with (\ref{2.2}). We disregard the cases
$\alpha S,_t = 0$ because they lead to the impossible condition $V,_t = 0$ --
see the penultimate paragraph of Sec. \ref{SFNmodels}. Then, (\ref{5.8}) is
compatible with (\ref{2.2}) only when $f = 0$, i.e. when the metric is
conformally flat. The conformally flat case belongs to the Stephani class
(\ref{2.4}), which will be discussed in part IV of this paper. However, in that
class the spherically symmetric case emerges among many others, and it will be
more convenient to discuss it here.\footnote{It may be verified that (\ref{5.6})
is indeed a first integral of (\ref{4.3}) modulo the RLP conditions. Hint: take
a derivative of (\ref{5.6}) along a null geodesic (defined by ${\cal D} / {\rm
d} r = (\dril t r) \pdril {} t + \pdril {} r$), then use (\ref{3.10}) and
(\ref{5.6}) to eliminate $\dril {^2t} {r^2}$ and $\dril \tau r$. The result will
be equal to (\ref{4.3}) provided that:
 $$
\frac {\tau} {{V,_t}^2} \dr t r \left(V,_{tt} V,_{tu} - V,_t V,_{ttu}\right)
\left[1 - \left(F V \dr t r\right)^2\right] = 0.
 $$
The expression in (\ ) vanishes when (\ref{5.7}) is fulfilled, the expression in
[\ ] vanishes in virtue of the null condition when the geodesic is radial. Thus,
(\ref{5.6}) is a first integral of (\ref{4.3}) when any of the RLP conditions is
fulfilled.}

When $f = 0$, (\ref{5.8}) substituted in (\ref{2.2}) gives $\alpha,_{uu} =
\beta,_{uu} = 0$, i.e.
\begin{equation}\label{5.9}
V = A_1 S + B_1 + \left(A_2 S + B_2\right) r^2.
\end{equation}
This is in the Stephani class (\ref{2.4}), with
\begin{eqnarray}\label{5.10}
&& \frac 1 {R(t)} = A_1 S + B_1, \qquad k(t) = 4 \frac {A_2 S + B_2} {A_1 S +
B_1},
\nonumber \\
&& x_0 = y_0 = z_0 = 0.
\end{eqnarray}
An FLRW limit results from (\ref{5.9}) when the zero-acceleration condition
(\ref{2.5}) is fulfilled, i.e. when $A_2 B_1 - A_1 B_2 = 0$. If $A_2 = 0$, then
in the FLRW limit $A_1 B_2 = 0$. But $A_1 = A_2 = 0$ is the prohibited case
$V,_t = 0$, while $B_2 = A_2 = 0$ is the $k = 0$ FLRW model. So, for a general
FLRW limit of (\ref{5.9}) we have $A_2 \neq 0$ and
\begin{equation}\label{5.11}
B_1 = A_1 B_2 / A_2.
\end{equation}
Then, in the FLRW limit
\begin{equation}\label{5.12}
V = \frac 1 {R(t)} \left(1 + \frac k 4 r^2\right),
\end{equation}
where
\begin{equation}\label{5.13}
R = \frac {A_2} {A_1 \left(A_2 S + B_2\right)}, \qquad k = \frac {4 A_2} {A_1}.
\end{equation}
Since $S$ is arbitrary and $\left(A_1, A_2\right)$ can have any signs, this
shows that the model defined by (\ref{5.9}) reproduces the whole FLRW class in
the zero-acceleration limit.

In the solution given by (\ref{5.9}) -- (\ref{5.10}) all null geodesics are
RLPs, since the metric obeys (\ref{5.7}) with no conditions imposed on the
vectors tangent to null geodesics.

Thus, the result is:

{\bf Corollary 1:} In a general spherically symmetric SFN model the only
repeatable light paths are radial null geodesics. In the subcase defined by
(\ref{5.9}) all null geodesics are RLPs. This subcase contains the whole FLRW
class, but is more general than FLRW because it has nonzero acceleration. It has
zero Weyl tensor, but is less general than the spherically symmetric limit of
the Stephani solution (\ref{2.4}); in the latter $k(t)$ and $R(t)$ are two
independent functions.

\part{\Large The plane symmetric models}

\section{The geodesic equations}\label{plangeo}

\setcounter{equation}{0}

The scheme of the calculation is here the same as in Sections \ref{sphergeo} --
\ref{spherlps}; only the explicit forms of the equations are different. Thus, we
will limit the explanation and the presentation  of intermediate expressions to
a necessary minimum.

Analogously to the spherical case we note that $z$ can be used as a parameter on
open intervals of each geodesic and use the $\dril {^2 z} {s^2}$ geodesic
equation to carry out the reparametrisation.\footnote{If $\dril z s = 0$ on an
open interval of the geodesic, then either (a) $\dril t s = 0$, which can hold
only on a spacelike geodesic, or (b) $V,_t = 0$, which is a prohibited case (see
the remark below (\ref{2.5})), or (c) $V,_{tz} = 0$, which is compatible with
(\ref{2.2}) only when $V,_t = 0$ or $f = 0$. The latter case is the conformally
flat model that will be dealt with in Sec. \ref{planelps}.} The geodesic
equations parametrised by $z$ are then:
\begin{eqnarray}
&& \dr {^2t} {z^2} - F^2 V,_t V,_{tz} \left(\dr t z\right)^3 + \left(\frac
{F,_t} F + \frac {V,_{t t}} {V,_t}\right) \left(\dr t z\right)^2 \nonumber \\
&&\ \ \ + 2 \frac {V,_{t z}} {V,_t} \dr t z = 0, \label{6.1} \\
&& \dr {^2 x} {z^2} - F^2 V,_t V,_{tz} \left(\dr t z\right)^2 \dr x z = 0,
\label{6.2} \\
&& \dr {^2 y} {z^2} - F^2 V,_t V,_{tz} \left(\dr t z\right)^2 \dr y z = 0.
\label{6.3}
\end{eqnarray}
Equations (\ref{6.2}) and (\ref{6.3}) show that $(y \dril x z - x \dril y z)$ is
a constant of the motion. This means that the projection of each geodesic on the
$(x, y)$ plane is a straight line.

In obtaining the redshift equations via the $\Delta$ operation defined in Sec.
\ref{spheredshift} we take $(\zeta, \psi) \df \Delta(x, y)$ and $(\xi, \eta) \df
(\dril {} z) (\zeta, \psi)$. Since we will not use the redshift equations in
full, we do not display them and proceed to the conditions for the RLPs.

\section{Repeatable light paths}\label{planelps}

\setcounter{equation}{0}

We substitute $\zeta = \psi = \xi = \eta = 0$ in the redshift equations and in
the null condition, and obtain the conditions for RLPs:
\begin{eqnarray}
&& \chi(t, z) \df \left[\left(2 F,_t V,_t + F V,_{tt}\right) V,_{tz} + F V,_t
V,_{ttz}\right] \dr t z \tau \nonumber \\
&&\ \ \ + 2 F V,_t V,_{tz} \dr \tau z, \label{7.1} \\
&& - F \chi \dr t z \dr x z = 0, \label{7.2} \\
&& - F \chi \dr t z \dr y z = 0, \label{7.3} \\
&& 2 F V,_t \dr t z \left[\left(F,_t V,_t + F V,_{tt}\right) \dr t z \tau + F
V,_t \dr \tau z\right] = 0.\ \ \ \ \ \  \label{7.4}
\end{eqnarray}
Discarding the impossible cases, there are two sets of solutions to (\ref{7.2})
-- (\ref{7.3}): (a) $\dril x z = \dril y z = 0$ -- the analogues of radial null
geodesics; (b) $\chi = 0$ -- the subset of the plane symmetric models in which
all null geodesics are RLPs. Using (\ref{7.4}) in $\chi = 0$ to eliminate $\dril
\tau z$, we obtain:
\begin{equation}\label{7.5}
V,_{tt} V,_{tz} - V,_t V,_{ttz} = 0,
\end{equation}
a general solution of which is
\begin{equation}\label{7.6}
V = \alpha(z) S(t) + \beta(z),
\end{equation}
where $\alpha$, $\beta$ and $S$ are arbitrary functions of their respective
arguments. Just as (\ref{5.8}), this is compatible with (\ref{2.2}) only when
the metric becomes conformally flat. Then $\alpha,_{zz} = \beta,_{zz} = 0$ and
\begin{equation}\label{7.7}
V = A_1 S + B_1 + \left(A_2 S + B_2\right) z.
\end{equation}
This case is not covered by the parametrisation of (\ref{2.4}), but can be
brought into the form of (\ref{2.4}) by the following special Haantjes
transformation:
\begin{eqnarray}\label{7.8}
(x, y, z) &=& \frac {\left(x', y', z' + C \left({x'}^2 + {y'}^2 +
{z'}^2\right)\right)} {\cal T}, \quad {\rm where} \nonumber \\
{\cal T} &\df& 1 + 2Cz' + C^2 \left({x'}^2 + {y'}^2 + {z'}^2\right),
\end{eqnarray}
where $C$ is an arbitrary constant, a group parameter of the transformation (see
Refs. \cite{Kras1989, PlKr2006} for hints on how to handle the calculations).
The result of (\ref{7.8}) is
\begin{eqnarray}\label{7.9}
&&V = C \left[C \left(A_1 S + B_1\right) + A_2 S + B_2\right]
\left({x'}^2 + {y'}^2 + {z'}^2\right) \nonumber \\
&&\ \ + \left[2 C \left(A_1 S + B_1\right) + A_2 S + B_2\right] z' + A_1 S +
B_1.\ \ \ \ \ \ \
\end{eqnarray}
For later reference we will need (\ref{7.9}) cast explicitly in the form
(\ref{2.4}). The functions appearing in (\ref{2.4}) are in our present case:
\begin{eqnarray}
\frac 1 R &=& - \frac {\left(A_2 S + B_2\right)^2} {4 \gamma}, \label{7.10} \\
\gamma &\df& C^2 \left(A_1 S + B_1\right) + C \left(A_2 S +
B_2\right), \label{7.11} \\
k &=& - \frac {16 \gamma^2} {\left(A_2 S + B_2\right)^2}, \label{7.12} \\
z_0 &=& \frac 1 {\gamma} \left[C \left(A_1 S + B_1\right) + \frac 1 2 \left(A_2
S + B_2\right)\right]. \ \ \ \ \label{7.13}
\end{eqnarray}
Equation (\ref{7.12}) suggests that $k < 0$ necessarily. However, the $k = 0$
case follows directly from (\ref{7.7}), it is $A_2 = B_2 =0$, and then $1 / R =
A_1 S + B_1$. Thus, the plane symmetric subcase of the Stephani model can
reproduce the $k \leq 0$ FLRW limits.

The general prescription for the FLRW limit can be found by substituting
(\ref{7.7}) in the zero-acceleration condition (\ref{2.4}); the result is
\begin{equation}\label{7.14}
A_1 B_2 - A_2 B_1 = 0.
\end{equation}

We will need the formulae for $R$ and $k$ parametrised by $z_0$ rather than by
$S$. From (\ref{7.13}) we find
\begin{equation}\label{7.15}
S = \frac {B_1 C + B_2 /2 - \left(B_1 C + B_2\right) C z_0} {\left(C A_1 +
A_2\right) C z_0 - C A_1 - A_2 / 2},
\end{equation}
and then from (\ref{7.12}) and (\ref{7.10})
\begin{eqnarray}
\frac 1 {2K} &=& \frac {\left(C A_1 + A_2\right) C z_0 - C A_1 - A_2 / 2} {C^2
\left(A_2 B_1 - A_1 B_2\right)},\ \ \  \label{7.16} \\
K &\df& \frac k {4R}, \label{7.17} \\
\frac 1 {2KR} &=& - \frac 1 2 {z_0}^2 + \frac {z_0} C - \frac 1 {2 C^2}.
\label{7.18}
\end{eqnarray}
Equations (\ref{7.16}) and (\ref{7.18}) will be needed to recognise the plane
symmetric subcase among the multitude of cases discussed in Appendix A.

Note that the form (\ref{7.16}) -- (\ref{7.18}) does not allow taking the FLRW
limit (\ref{7.14}). This is logical, since in this limit $z_0$ becomes constant
while $K$ and $R$ do not, so $z_0$ cannot be used as a parameter.

We thus have:

{\bf Corollary 2:} In a general plane symmetric SFN model represented as in
(\ref{2.1}) -- (\ref{2.3}) the only repeatable light paths are null geodesics on
which $x$ and $y$ are constant. In the subcase defined by (\ref{7.7}), which
includes the $k \leq 0$ FLRW subclass, all null geodesics are RLPs. This subcase
is conformally flat, but less general than the conformally flat limit of the
plane symmetric case.

The condition $k \leq 0$ is consistent with what is known about the relation
between plane symmetric models and their FLRW limits \cite{Kras1997}.

\part{\Large The hyperbolically symmetric models}

\section{The geodesic equations}\label{hypergeo}

\setcounter{equation}{0}

There is a certain complication in discussing this case: either the metric can
be similar to (\ref{2.1}) or the equation similar to (\ref{2.2}), but not both
things at once. We choose to make the metric similar. We use eq. (2.1) of Ref.
\cite{Kras1989}, rename the functions by $(3\theta, Y, b) \df (1/F, V, f)$ and
transform the $r$-coordinate used there by $r = \exp(r')$. Dropping the prime,
the metric then becomes:
\begin{equation}
{\rm d} s^2 = \left(\frac {F V,_t} V\right)^2 {\rm d} t^2 - \frac 1 {V^2}
\left({\rm d} r^2 + {\rm d} \vartheta^2 + \sinh^2 \vartheta {\rm d}
\varphi^2\right), \label{8.1}
\end{equation}
and the function $V(t, r)$ must obey:
\begin{equation}
V,_{rr} = f(r) V^2 - V. \label{8.2}
\end{equation}
See \cite{Kras1989} for the transformation from (\ref{8.1}) to (\ref{2.1}) --
(\ref{2.3}).

Just as in Sec. \ref{plangeo} we will now skip most of the explanation and of
the intermediate expressions because the calculations exactly parallel those for
the spherically symmetric models presented in Sections \ref{sphergeo} --
\ref{spherlps}.

Using the $\dril {^2 r} {s^2}$ geodesic equation we change the parameter to the
non-affine $r$ and obtain:
\begin{eqnarray}
\dr {^2 t} {r^2} &+& 2 \frac {V,_{t r}} {V,_t} \dr t r + \left(\frac {F,_t} F +
\frac {V,_{t t}} {V,_t}\right) \left(\dr t r\right)^2 \nonumber \\
&-& F^2 V,_t V,_{t r} \left(\dr t r\right)^3 = 0, \label{8.3} \\
\dr {^2 \vartheta} {r^2} &-& F^2 V,_t V,_{t r} \dr \vartheta r \left(\dr t
r\right)^2 - \cosh \vartheta \sinh \vartheta \left(\dr \varphi r\right)^2
\nonumber \\
&&\ \ \ \ \ \ \ \ \ \ \ \ \ \ = 0, \label{8.4} \\
\dr {^2 \varphi} {r^2} &-& F^2 V,_t V,_{t r} \dr \varphi r \left(\dr t
r\right)^2+ 2 \coth \vartheta \dr \vartheta r \dr \varphi r = 0. \nonumber \\
\label{8.5}
\end{eqnarray}

We skip through the general redshift equations and proceed to the RLP equations,
where this time $(\zeta, \psi) \df \Delta(\vartheta, \varphi)$ and $(\xi, \eta)
\df (\dril {} z) (\zeta, \psi)$.

\section{Repeatable light paths}\label{hyperlps}

\setcounter{equation}{0}

When $\zeta = \psi = \xi = \eta = 0$, the redshift equations become:
\begin{eqnarray}
&& \chi(t, r) \df \left[\left(2 F,_t V,_t + F V,_{tt}\right) V,_{tr} + F V,_t
V,_{ttr}\right] \dr t r \tau \nonumber \\
&&\ \ \ \ \ \ \ \ \ \ \ \ + 2 F V,_t V,_{tr} \dr \tau r,  \label{9.1} \\
&&- F \chi \dr t r \dr \vartheta r = 0, \label{9.2} \\
&&- F \chi \dr t r \dr \varphi r = 0, \label{9.3} \\
&& 2 F V,_t \left[\left(F,_t V,_t + F V,_{tt}\right) \dr t r \tau + F V,_t \dr
\tau r\right] \dr t r = 0. \nonumber \\
 \label{9.4}
\end{eqnarray}
As before, (\ref{9.4}) implies:
\begin{equation}\label{9.5}
\dr \tau r = - \left(\frac {F,_t} F + \frac {V,_{tt}} {V,_t}\right) \dr t r
\tau.
\end{equation}
Equations (\ref{9.2}) -- (\ref{9.3}) tell us that either $\dril \vartheta r =
\dril \varphi r = 0$ or $\chi = 0$. The first case defines a pseudo-radial null
geodesic. The other possibility is $\chi = 0$. Using (\ref{9.5}) we get
\begin{equation}\label{9.6}
V,_{tt} V,_{tr} - V,_t V,_{ttr} = 0.
\end{equation}
A general solution of this equation is
\begin{equation}\label{9.7}
V = \alpha(r) S(t) + \beta(r),
\end{equation}
where $\alpha$, $\beta$ and $S$ are arbitrary functions of their respective
arguments. As before, this is compatible with (\ref{8.2}) only when $f = 0$ and
the solution becomes conformally flat. Then, from (\ref{8.2}), $V$ must obey,
$V,_{rr} + V = 0$, and the solution is
\begin{equation}\label{9.8}
V = \left(A_1 S + B_1\right) \sin r + \left(A_2 S + B_2\right) \cos r.
\end{equation}
This is transformed to the standard parametrisation (\ref{2.4}) by the following
chain of transformations.

We first transform the 2-dimensional metric $({\rm d} \vartheta^2 + \sinh^2
\vartheta {\rm d} \varphi^2)$ into $({\rm d} \tau^2 + {\rm e}^{2\tau} {\rm d}
z^2)$. An explicit prescription for this is given in Appendix A to Ref.
\cite{Kras1989}. This does not affect $V$ because $\vartheta$ and $\varphi$ are
not present in it. Then we carry out the transformation
\begin{equation}\label{9.9}
r = \arcsin \frac y {\sqrt{x^2 + y^2}}, \qquad \tau = - \frac 1 2 \ln \left(x^2
+ y^2\right).
\end{equation}
The metric then becomes:
\begin{equation}\label{9.10}
{\rm d} s^2 = \left(\frac {FW,_t} W\right)^2 {\rm d} t^2 - \frac {{\rm d} x^2 +
{\rm d} y^2 + {\rm d} z^2} {W^2},
\end{equation}
where
\begin{equation}\label{9.11}
W \df \sqrt{x^2 + y^2} V = \left(A_1 S + B_1\right) y + \left(A_2 S + B_2\right)
x.
\end{equation}
The metric (\ref{9.10}) -- (\ref{9.11}) is within the Stephani class, as can be
seen by carrying out the following Haantjes transformation:
\begin{equation}\label{9.12}
(x, y, z) = \frac {\left(x' + C \left({x'}^2 + {y'}^2 + {z'}^2\right), y',
z'\right)} {1 + 2Cx' + C^2 \left({x'}^2 + {y'}^2 + {z'}^2\right)}.
\end{equation}
After this, the metric acquires the form (\ref{2.1}) with $V$ replaced by
\begin{eqnarray}\label{9.13}
W_1 &=& C \left(A_2 S + B_2\right) \left({x'}^2 + {y'}^2 + {z'}^2\right)
\nonumber
\\
&&\ \ \ + \left(A_2 S + B_2\right) x' + \left(A_1 S + B_1\right) y'.\ \ \ \
\end{eqnarray}
When $W_1$ is cast in the form of (\ref{2.4}), we obtain
\begin{eqnarray}
&& x_0 = - \frac 1 2, \label{9.14} \\
&& y_0 = - \frac 1 2 \frac {A_1 S + B_1} {A_2 S + B_2}, \label{9.15} \\
&& K \df \frac k {4R} = C \left(A_2 S + B_2\right), \label{9.16} \\
&& \frac 1 R = - \frac C 4 \times \frac {\left(A_1 S + B_1\right)^2 + \left(A_2
S + B_2\right)^2} {A_2 S + B_2}.\ \ \ \ \ \ \  \label{9.17}
\end{eqnarray}
The constant $x_0$ can be transformed to $x_0 = 0$.

The FLRW limit follows when $A_1 B_2 - A_2 B_1 = 0$. From (\ref{9.16}) --
(\ref{9.17}) we find
\begin{equation}\label{9.18}
k = - \frac {16 \left(A_2 S + B_2\right)^2} {\left(A_1 S + B_1\right)^2 +
\left(A_2 S + B_2\right)^2},
\end{equation}
which shows that $k < 0$ necessarily. The $V$ given by (\ref{9.8}) cannot be
made independent of $r$, so the case $k = 0$ is not contained in this model.

For later reference we will need the formulae for $K$ and $R$ parametrised by
$y_0$ rather than by $S$. We have
\begin{eqnarray}
&& S = - \frac {B_1 + 2 B_2 y_0} {A_1 + 2 A_2 y_0}, \label{9.19} \\
&& \frac 1 {2K} = \frac {A_1 + 2 A_2 y_0} {2C \left(A_1 B_2 - A_2 B_1\right)},
\label{9.20} \\
&& \frac 1 {2 KR} = - \frac 1 {2C} {y_0}^2 - \frac 1 {8C}. \label{9.21}
\end{eqnarray}
Similarly to the plane symmetric model, in this form the FLRW limit $A_1 B_2 -
A_2 B_1 = 0$ cannot be taken, for the same reason: in the FLRW models $y_0$ is
constant, while $K$ and $R$ are not, so $y_0$ cannot be used as a parameter.

Thus we have:

{\bf Corollary 3:} In a general hyperbolically symmetric SFN model the only
repeatable light paths are the analogues of radial null geodesics, on which
$\vartheta$ and $\varphi$ in (\ref{8.1}) are constant. In the subcase defined by
(\ref{9.10}) -- (\ref{9.11}), which includes the FLRW models with $k < 0$, all
null geodesics are RLPs. This subcase is conformally flat, but less general than
the conformally flat limit of (\ref{8.1}).

\part{\Large The Stephani model}\label{Stepmodel}

\section{The geodesic equations}\label{Stepgeo}

\setcounter{equation}{0}

The metric of this model is (\ref{2.1}), with $V$ given by (\ref{2.4}). We
introduce the abbreviation:
\begin{equation}
D \df F(t) V,_t / V. \label{10.1}
\end{equation}
The geodesic equations in the affine parametrisation, with the null condition
already incorporated, are:
\begin{eqnarray}
\dr {^2t} {s^2} &+& 2 \left(\frac {D,_x} D \dr x s + \frac {D,_y} D \dr y s +
\frac {D,_z} D \dr z s\right) \dr t s \nonumber \\
&+& \left(\frac {D,_t} D - \frac {V,_t} V\right) \left(\dr t
s\right)^2 = 0, \label{10.2} \\
\dr {^2x} {s^2} &+& F^2 V,_t V,_{tx} \left(\dr t s\right)^2 - 2
\frac {V,_x} V \left(\dr x s\right)^2 \nonumber \\
&-& 2 \left(\frac {V,_t} V \dr t s + \frac {V,_y} V \dr y s +
\frac {V,_z} V \dr z s\right) \dr x s = 0,\ \ \ \ \ \  \label{10.3} \\
\dr {^2y} {s^2} &+& F^2 V,_t V,_{ty} \left(\dr t s\right)^2 - 2
\frac {V,_y} V \left(\dr y s\right)^2 \nonumber \\
&-& 2 \left(\frac {V,_t} V \dr t s + \frac {V,_x} V \dr x s +
\frac {V,_z} V \dr z s\right) \dr y s = 0,\ \ \ \ \ \  \label{10.4} \\
\dr {^2z} {s^2} &+& F^2 V,_t V,_{tz} \left(\dr t s\right)^2 - 2
\frac {V,_z} V \left(\dr z s\right)^2 \nonumber \\
&-& 2 \left(\frac {V,_t} V \dr t s + \frac {V,_x} V \dr x s + \frac {V,_y} V \dr
y s\right) \dr z s = 0.\ \ \ \ \ \  \label{10.5}
\end{eqnarray}
Note that if any of $\dril {x^i} s$, $i = 1, 2, 3$ (where $(x^1, x^2, x^3)$ $\df
(x, y, z)$) is zero along an open interval of a null geodesic, then (\ref{10.2})
-- (\ref{10.5}) imply either $\dril t s = 0$ in the same interval (which is
impossible on a null curve) or $V,_{t i} = 0$ for the respective $i$. This
second possibility must be investigated. Suppose first that
\begin{equation}\label{10.6}
V,_{t x} = V,_{t y} = V,_{t z} = 0.
\end{equation}
This means:
\begin{equation}\label{10.7}
\left(\frac k {2R}\right),_t = x_{0,t} = y_{0,t} = z_{0,t} = 0.
\end{equation}
The metric is then spherically symmetric and the constants $(x_0, y_0, z_0)$ can
be set equal to zero by coordinate transformations. This case was dealt with at
the end of Sec. \ref{spherlps}, so we disregard it here.

Hence, we may assume that at least one of the quantities in (\ref{10.6}) is
nonzero. Since a general Stephani metric does not change its form under any
permutation of the $(x, y, z)$ coordinates (accompanied by a suitable renaming
of the $(x_0, y_0, z_0)$ functions), we may assume without loss of generality
that $ V,_{t z} \neq 0$, and then (\ref{10.5}) shows that $\dril z s$ cannot be
zero over an open interval of a null geodesic. Consequently, $z$ can be chosen
as a (non-affine) parameter, with an analogous cautionary remark to the one
given above (\ref{3.8}).

Consequently, using (\ref{10.5}) and (\ref{3.8}) for $\dril {^2z} {s^2}$ we
change the parameter to $z$ in (\ref{10.2}) -- (\ref{10.4}) and obtain:
\begin{eqnarray}
\dr {^2t} {z^2} &-& F^2 V,_t V,_{tz} \left(\dr t z\right)^3 + \left(\frac {F,_t}
F + \frac {V,_{tt}} {V,_t}\right) \left(\dr t z\right)^2 \nonumber \\
&+& 2 \left(\frac {V,_{tx}} {V,_t} \dr x z + \frac {V,_{ty}}
{V,_t} \dr y z + \frac {V,_{tz}} {V,_t}\right) \dr t z = 0, \label{10.8} \\
\dr {^2x} {z^2} &-& F^2 V,_t V,_{tz} \dr x z \left(\dr t z\right)^2 +
F^2 V,_t V,_{tx} \left(\dr t z\right)^2 = 0, \nonumber \\
\label{10.9} \\
\dr {^2y} {z^2} &-& F^2 V,_t V,_{tz} \dr y z \left(\dr t z\right)^2 +
F^2 V,_t V,_{ty} \left(\dr t z\right)^2 = 0, \nonumber \\
\label{10.10}
\end{eqnarray}

\section{Repeatable light paths}\label{Steprlps}

\setcounter{equation}{0}

In this case we skip the redshift equations because they are complicated and
voluminous, while we are not going to make any direct use of them. In this
section, the meaning of the symbols in the $\Delta$ operation is
\begin{equation}\label{11.1}
\Delta(t, x, y, \dril x z, \dril y z) \df (\tau, \zeta, \psi, \xi, \eta).
\end{equation}
We proceed to the RLP conditions that are obtained by applying the $\Delta$
operation to (\ref{10.9}) -- (\ref{10.10}) and immediately assuming $\zeta =
\psi = \xi = \eta = 0$.\footnote{Similarly to what footnote \#1 says, a quick
way to obtain the equations that follow is to take the differentials of
(\ref{10.8}) -- (\ref{10.10}) at constant $(x, y, z)$ and replace ${\rm d} t$ by
$\tau$.} We do not consider the result of $\Delta$ acting on (\ref{10.8})
because it is the equation for $\tau$ that will define the redshift propagation
along the emergent RLP, and it does not lead to any limitation on the metric.

The null condition in the $z$-parametrization is:
\begin{equation}\label{11.2}
F^2 {V,_t}^2 \left(\dr t z\right)^2 = \left(\dr x z\right)^2 + \left(\dr y
z\right)^2 + 1,
\end{equation}
and the RLP condition resulting from it is:
\begin{equation}\label{11.3}
\left(F,_t V,_t + F V,_{tt}\right) \dr t z \tau + FV,_t \dr \tau z = 0
\end{equation}
(we ignore the cases $FV,_t = 0$ -- prohibited in (\ref{2.1}), and $\dril t z =
0$ -- which defines a spacelike curve).

We denote:
\begin{equation}\label{11.4}
H_i \df \frac {V,_{ti}} {V,_t}, \qquad G_i \df H_{i,t}, \qquad i = 1, 2, 3,
\end{equation}
where $(x^1, x^2, x^3) \df (x, y, z)$. We find $\dril \tau z$ from (\ref{11.3})
and use it in the RLP conditions resulting from (\ref{10.9}) -- (\ref{10.10}) in
the way described above. The result is:
\begin{eqnarray}
&& G_x  - G_z \dr x z = 0, \label{11.5} \\
\nonumber \\
&& G_y - G_z \dr y z = 0. \label{11.6}
\end{eqnarray}
Since the analysis of (\ref{11.5}) -- (\ref{11.6}) is complicated, we state here
only the results, and present the calculations in Appendix \ref{genSteprlp}. The
RLPs defined by (\ref{11.5}) -- (\ref{11.6}) exist in the following cases:

{\underline {(1) (Case 1.1.1.1.2.1.1 of Appendix \ref{genSteprlp})}}

When $y_0$ is defined by (\ref{a.23}), $K \df k / (4R)$ by (\ref{a.24}) and $R$
by (\ref{a.39}), some of the null geodesics are RLPs, and they are solutions of
(\ref{a.52}), with ${\cal F}_1$ and ${\cal G}_1$ given by (\ref{a.40}) --
(\ref{a.42}) and (\ref{a.50}) -- (\ref{a.51}). It is not known whether this
subcase of the Stephani solution admits an FLRW limit. If it does, then only
after a reparametrisation. As noted for the plane and hyperbolically symmetric
cases, when $k$ and $R$ are parametrised by $x_0$, $y_0$ or $z_0$, the FLRW
limit cannot be calculated.

Several subcases of this spacetime appear separately in Appendix
\ref{genSteprlp}, but they are not listed here.

{\underline {(2) (Case 1.1.1.2.1.2.1 of Appendix \ref{genSteprlp})}}

When $y_0 = D_1 x_0$, $z_0 = C_3 x_0$, and $R$ is determined by (\ref{a.64}),
but $K(t) = k / (4R)$ is arbitrary, again some of the null geodesics are RLPs.
They are determined by (\ref{a.52}), where ${\cal F}_1$ and ${\cal G}_1$ are
given by (\ref{a.66}) -- (\ref{a.67}).

Also here, several subcases appear separately, but they do not admit more RLPs.

{\underline {(3) (Case 1.1.2.2.2 of Appendix \ref{genSteprlp})}}

This is a copy of the subcase of the spherically symmetric Stephani solution
discussed in Sec. \ref{spherlps}. All of its null geodesics are RLPs, and it
contains the FLRW limit in full generality. It is not identical to FLRW because
in general it has nonzero acceleration.

{\underline {(4) (Case 1.2.1.2 of Appendix \ref{genSteprlp})}}

Then $x_0 = C_1 z_0$, $y_0 = C_3 z_0$, $R(t)$ is arbitrary, $V$ and $k$ are
determined by (\ref{a.111}) and (\ref{a.112}). This is an axially symmetric
subcase of the Stephani model, and its RLPs are determined by (\ref{a.113}). All
the RLPs intersect the symmetry axis $x = y = 0$ in each space of constant $t$.

{\underline {(5) (Case 2.1 of Appendix \ref{genSteprlp})}}

Then $y_0 = C_2 x_0$, $z_0 = C_3 x_0$, $K = k / (4R)$ is determined by
(\ref{a.137}) and $R$ is determined by (\ref{a.141}). As indicated there, $C_2 =
C_3 = 0$ may be achieved by a coordinate transformation, and then the model is
seen to be axially symmetric. All of its null geodesics are RLPs. It is less
general than the previous one (see under (4) above) because the former has
$R(t)$ arbitrary, while the current one has $R$ determined by (\ref{a.141}).

\section{Summary}\label{summ}

\setcounter{equation}{0}

The existence of repeatable light paths (RLPs) was investigated for the
cosmological models found by Barnes \cite{Barn1973} and Stephani
\cite{Step1967}. They are called shearfree normal (SFN) because the perfect
fluid source in the Einstein equations moves with zero shear and zero rotation.
The Barnes models are either spherically symmetric (SS) or plane symmetric (PS)
or hyperbolically symmetric (HS). In general, in each of these classes only
those null geodesics are RLPs that are orthogonal to the symmetry orbits (the
radial ones in the SS case). However, each one contains a subclass in which all
null geodesics are RLPs. This subclass is in each case more general than FLRW.
In the SS case, this special subclass has the metric (\ref{2.4}) with $V$ given
by (\ref{5.9}) and (\ref{5.10}), and contains the whole FLRW family. In the PS
case, the special subclass has $V$ given by (\ref{7.9}) -- (\ref{7.13}) and
contains only those FLRW models for which the curvature index $k \leq 0$. In the
HS case, the special subclass is given by (\ref{9.13}) -- (\ref{9.17}) and
contains only the $k < 0$ FLRW models. All these special subclasses are
conformally flat, but less general than the corresponding conformally flat
limits of the SS, PS and HS cases.

For the Stephani models the situation is summarised in Sec. \ref{Steprlps}. In
general, no RLPs exist. The conditions of existence of RLPs, (\ref{11.5}) --
(\ref{11.6}), put limitations on the Stephani metric. There exist subcases in
which some of the null geodesics are RLPs, for example the axially symmetric
subcase given by (\ref{a.111}) -- (\ref{a.112}). There exists also a subclass in
which all null geodesics are RLPs, it is axially symmetric as well.

This study, as explained in the introduction, was motivated by the question:
what is the geometrical condition for the existence of RLPs; is it vanishing
shear in the flow of the cosmic medium, as suggested by the result of Ref.
\cite{KrBo2011}, or rather a high symmetry of the spacetime? The solutions of
Einstein's equations investigated in the present paper all have zero shear, and
yet their null geodesics are RLPs only in special situations. Thus, the key to
the RLP property is a symmetry of the spacetime rather than vanishing shear.

\appendix

\section{RLPs obeying (\ref{11.5}) -- (\ref{11.6})}\label{genSteprlp}

\setcounter{equation}{0}

The calculations in this Appendix are trivial in principle. The reason why they
are presented in some detail is that the various separate subcases form a very
complicated binary tree that would be difficult to duplicate without a
guidebook.

\medskip

{\bf \underline {Case 1: The general case: $G_z \neq 0$}}

\medskip

Then (\ref{11.5}) -- (\ref{11.6}) become
\begin{equation}\label{a.1}
\dr x z = \frac {G_x} {G_z}, \qquad \dr y z = \frac {G_y} {G_z}.
\end{equation}
We calculate the derivatives of $\dril x z$ and $\dril y z$ along a null
geodesic by the rule
 $$
\frac {\rm D} {{\rm d} z} = \dr t z \pdr {} t + \dr x z \pdr {} x + \dr y z \pdr
{} y +  \pdr {} z
 $$
and substitute the results in (\ref{10.9}) -- (\ref{10.10}). Then we use
(\ref{a.1}) and (\ref{11.2}) to eliminate $\dril x z$, $\dril y z$ and $(\dril t
z)^2$. The resulting equations are
\begin{eqnarray}
&& G_z \left(G_z G_{x,t} - G_x G_{z,t}\right) \dr t z + G_x \left(G_z G_{x,x} -
G_x G_{z,x}\right) \nonumber \\
&& + G_y \left(G_z G_{x,y} - G_x G_{z,y}\right) + G_z \left(G_z G_{x,z} - G_x
G_{z,z}\right) \nonumber \\
&& + \left(H_x G_z - H_z G_x\right) \left({G_x}^2 + {G_y}^2 + {G_z}^2\right) =
0, \label{a.2} \\
&& G_z \left(G_z G_{y,t} - G_y G_{z,t}\right) \dr t z + G_x \left(G_z G_{y,x} -
G_y G_{z,x}\right) \nonumber \\
&& + G_y \left(G_z G_{y,y} - G_y G_{z,y}\right) + G_z \left(G_z G_{y,z} - G_y
G_{z,z}\right) \nonumber \\
&& + \left(H_y G_z - H_z G_y\right) \left({G_x}^2 + {G_y}^2 + {G_z}^2\right) =
0. \label{a.3}
\end{eqnarray}
After (\ref{11.4}) are substituted for $G_i$ and $H_i$, the terms free of $\dril
t z$ sum up to zero in each of (\ref{a.2}) -- (\ref{a.3}), so both coefficients
of $\dril t z$ must be zero, too. This implies:
\begin{equation}\label{a.4}
\left(\frac {G_x} {G_z}\right),_t = \left(\frac {G_y} {G_z}\right),_t = 0.
\end{equation}
The integrals of these are:
\begin{equation}\label{a.5}
G_x = {\cal F}_1(x, y, z) G_z, \qquad G_y = {\cal G}_1(x, y, z) G_z,
\end{equation}
where ${\cal F}_i$ and ${\cal G}_i$, $i = 1, 2, 3$ are arbitrary functions.
Recalling (\ref{11.4}), these are integrated again with the result:
\begin{eqnarray}\label{a.6}
&& H_x = {\cal F}_1 H_z + {\cal F}_2(x, y, z), \nonumber \\
&& H_y = {\cal G}_1 H_z + {\cal G}_2(x, y, z).
\end{eqnarray}
Finally, recalling the definitions of $H_x$ and $H_y$ from (\ref{11.4}), eqs.
(\ref{a.6}) are integrated with the result:
\begin{eqnarray}
&& V,_x = {\cal F}_1 V,_z + {\cal F}_2 V + {\cal F}_3(x, y, z), \label{a.7} \\
&& V,_y = {\cal G}_1 V,_z + {\cal G}_2 V + {\cal G}_3(x, y, z). \label{a.8}
\end{eqnarray}
We introduce the following conventions:

$\bullet$ $C_i, D_i, c_i, d_i, E_i, i = 1,2, \dots$ will denote arbitrary
constants,

$\bullet$ ${\cal F}_j, {\cal G}_j, j = 4, 5, \dots$ will denote arbitrary
functions of spatial coordinates, not necessarily of all of them.

\noindent Since the alternatives considered below will be in most cases mutually
exclusive, we will re-use the same names of constants and functions with
different meanings.

In every case we will follow the same scheme of reasoning. Our initial equation
(IE) will be (\ref{a.7}) or (\ref{a.8}), usually multiplied by some factor
($1/2K$ or ${{\cal F}_2}^{-1}$ or ${{\cal G}_2}^{-1}$). Then, the following
operations will be executed on IE, each one followed by conclusions:

1) Differentiate IE by $y$ and $t$;

2) Differentiate IE by $y$ alone;

3) Differentiate IE by $x$ and $t$;

4) Differentiate IE by $x$ alone;

5) Differentiate IE by $z$ and $t$;

6) Differentiate IE by $z$ alone.

\noindent At this stage, $({\cal F}_i, {\cal G}_i), i = 1, 2, 3, K$ and $R$ will
be defined, and the corresponding Stephani model admitting RLPs (if any) will be
identified. The functions $({\cal F}_1, {\cal G}_1)$, substituted in
(\ref{a.1}), will define the RLPs.

We will present in detail the whole 6-step procedure only for the first case
considered.

It will turn out in several places along the way that the function $x_0(t)$ is
in fact constant. In those cases, we will assume that $x_0 = 0$ because this
result can be achieved by the coordinate transformation $x = x' + x_0$. The same
is true for the pairs $(y_0, y)$ and $(z_0, z)$.

\medskip

{\bf \underline {Case 1.1: ${\cal F}_2 \neq 0$}}

\medskip

Then (\ref{a.7}) can be written as:
\begin{eqnarray}\label{a.9}
&& \frac {x - x_0} {{\cal F}_2} = \left(z - z_0\right) \frac {{\cal F}_1} {{\cal
F}_2} + \frac 1 {2KR} \\
&& + \frac 1 2 \left[\left(x - x_0\right)^2 + \left(y - y_0\right)^2 + \left(z -
z_0\right)^2\right] + \frac 1 {2K} \frac {{\cal F}_3} {{\cal F}_2}, \nonumber
\end{eqnarray}
where we have introduced the symbol:
\begin{equation}\label{a.10}
K \df \frac k {4 R}.
\end{equation}
In writing (\ref{a.9}) we assumed $k \neq 0$ because $k = 0$ is the spatially
flat FLRW model, in which we know that all null geodesics are RLPs. Taking the
second derivative of (\ref{a.9}) by $y$ and $t$ we obtain:
\begin{equation}\label{a.11}
x_{0,t} \left(\frac 1 {{\cal F}_2}\right),_y = z_{0,t} \left(\frac {{\cal F}_1}
{{\cal F}_2}\right),_y + y_{0,t} - \left(\frac 1 {2K}\right),_t \left(\frac
{{\cal F}_3} {{\cal F}_2}\right),_y.
\end{equation}

\medskip

{\bf \underline {Case 1.1.1: $x_{0,t} \neq 0$}}

\medskip

Then we divide (\ref{a.11}) by $x_{0,t}$ and differentiate the result by $t$,
obtaining:
\begin{eqnarray}\label{a.12}
&& \left(\frac {z_{0,t}} {x_{0,t}}\right),_t \left(\frac {{\cal F}_1} {{\cal
F}_2}\right),_y + \left(\frac {y_{0,t}} {x_{0,t}}\right),_t \nonumber \\
&& - \left[\frac 1 {x_{0,t}} \left(\frac 1 {2K}\right),_t\right],_t \left(\frac
{{\cal F}_3} {{\cal F}_2}\right),_y = 0.
\end{eqnarray}

\medskip

{\bf \underline {Case 1.1.1.1: $\left(z_{0,t} / x_{0,t}\right),_t \neq 0$}}

\medskip

Then we divide (\ref{a.12}) by $\left(z_{0,t} / x_{0,t}\right),_t$ and
differentiate the result by $t$. We get:
\begin{equation}\label{a.13}
\left[\left(\frac {y_{0,t}} {x_{0,t}}\right),_t\right/\left.\left(\frac
{z_{0,t}} {x_{0,t}}\right),_t\right],_t - \chi(t) \left(\frac {{\cal F}_3}
{{\cal F}_2}\right),_y = 0,
\end{equation}
where:
\begin{equation}\label{a.14}
\chi(t) \df \left\{\left[\frac 1 {x_{0,t}} \left(\frac 1
{2K}\right),_t\right],_t \right/\left.\left(\frac {z_{0,t}}
{x_{0,t}}\right),_t\right\},_t.
\end{equation}

\medskip

{\underline {\bf Case 1.1.1.1.1: $\chi(t) \neq 0$}}

\medskip

Then we get:
\begin{eqnarray}\label{a.15}
&& \frac 1 {\chi} \left[\left(\frac {y_{0,t}}
{x_{0,t}}\right),_t\right/\left.\left(\frac {z_{0,t}}
{x_{0,t}}\right),_t\right],_t = \left(\frac {{\cal F}_3} {{\cal F}_2}\right),_y
\nonumber \\
&& \ \ \ \ \ = C_1 = {\rm constant},
\end{eqnarray}
both expressions being constant because the first one depends only on $t$, while
the second one depends only on $(x, y, z)$. Integrating both equations we
obtain:
\begin{eqnarray}
&& \frac {C_1} {2K} = D_1 x_0 + y_0 + D_3 z_0 + D_4, \label{a.16} \\
&& {\cal F}_3 = \left(C_1 y + {\cal F}_4(x, z)\right) {\cal F}_2. \label{a.17}
\end{eqnarray}
In principle, we would have to consider the cases $C_1 \neq 0$ and $C_1 = 0$
separately. However, they lead to the same result. When $C_1 \neq 0$,
(\ref{a.16}) determines $1/(2K)$. Then we substitute (\ref{a.16}) and
(\ref{a.17}) in (\ref{a.12}) and obtain
\begin{equation}\label{a.18}
\left(\frac {z_{0,t}} {x_{0,t}}\right),_t \left[\left(\frac {{\cal F}_1} {{\cal
F}_2}\right),_y - D_3\right] = 0.
\end{equation}
Since in Case 1.1.1.1 the first factor is nonzero, we have:
\begin{equation}\label{a.19}
{\cal F}_1 = \left(D_3 y + {\cal F}_5(x, z)\right) {\cal F}_2.
\end{equation}
When we substitute (\ref{a.17}) and (\ref{a.19}) in (\ref{a.11}) we obtain:
\begin{equation}\label{a.20}
x_{0,t} \left[\left(\frac 1 {{\cal F}_2}\right),_y + D_1\right] = 0.
\end{equation}
Since $x_{0,t} \neq 0$ in our current Case 1.1.1, we have:
\begin{equation}\label{a.21}
\frac 1 {{\cal F}_2} = - D_1 y + {\cal F}_6(x, z).
\end{equation}
Using (\ref{a.21}), (\ref{a.19}) and (\ref{a.17}) in (\ref{a.9}) we obtain:
\begin{eqnarray}
&& x \left(- D_1 y + {\cal F}_6\right) - x_0 {\cal F}_6 = z \left(D_3 y + {\cal
F}_5\right) - z_0 {\cal F}_5 \nonumber \\
&& + \frac 1 {2KR} + \frac 1 2 \left[\left(x - x_0\right)^2 + y^2 + {y_0}^2 +
\left(z - z_0\right)^2\right] \nonumber \\
&& + \frac 1 {2K} {\cal F}_4 + D_4 y. \label{a.22}
\end{eqnarray}
The derivative of this by $y$ now gives $- D_1 x = D_3z + y + D_4$, which is a
clear contradiction.

When $C_1 = 0$, $K$ remains undetermined, and instead (\ref{a.15}) determines
$y_0 = - D_1 x_0 - D_3 z_0$. Equations (\ref{a.19}) and (\ref{a.21}) still
follow and (\ref{a.22}) results with $D_4 = 0$, leading to the same
contradiction.

Case 1.1.1.1.1 thus turned out to be empty, and we go back to (\ref{a.13}) to
consider:

\medskip

{\underline {\bf Case 1.1.1.1.2: $\chi(t) = 0$}}

\medskip

Then (\ref{a.13}) implies $\left[\left(y_{0,t} /
x_{0,t}\right),_t\right/\left.\left(z_{0,t} / x_{0,t}\right),_t\right],_t = 0$,
which is integrated with the result:
\begin{equation}\label{a.23}
y_0 = C_1 x_0 + C_3 z_0
\end{equation}
(the additive constant was set to zero by a transformation $y = y' +$ constant),
and $\chi(t) = 0$ implies, via (\ref{a.14})
\begin{equation}\label{a.24}
\frac 1 {2K} = D_1 x_0 + D_3 z_0 + D_4.
\end{equation}
We substitute (\ref{a.23}) -- (\ref{a.24}) in (\ref{a.12}) and obtain
\begin{equation}\label{a.25}
\left(\frac {z_{0,t}} {x_{0,t}}\right),_t \left[\left(\frac {{\cal F}_1} {{\cal
F}_2}\right),_y - D_3 \left(\frac {{\cal F}_3} {{\cal F}_2}\right),_y +
C_3\right] = 0.
\end{equation}
Since we are still in Case 1.1.1.1, where $\left(z_{0,t} / x_{0,t}\right),_t
\neq 0$, the above leads to
\begin{equation}\label{a.26}
{\cal F}_1 = D_3 {\cal F}_3 + \left({\cal F}_4(x,z) - C_3 y\right) {\cal F}_2.
\end{equation}
Substituting this and (\ref{a.23}) -- (\ref{a.24}) in (\ref{a.11}) we get
\begin{equation}\label{a.27}
x_{0,t} \left[\left(\frac 1 {{\cal F}_2}\right),_y - C_1 + D_1 \left(\frac
{{\cal F}_3} {{\cal F}_2}\right),_y\right] = 0.
\end{equation}
In Case 1.1.1, where $x_{0,t} \neq 0$, the above implies
\begin{equation}\label{a.28}
D_1 {\cal F}_3 = \left(C_1 y + {\cal F}_5(x,z)\right) {\cal F}_2 - 1.
\end{equation}

\medskip

{\underline {\bf Case 1.1.1.1.2.1: $D_1 \neq 0$}}

\medskip

Then (\ref{a.28}) determines ${\cal F}_3$. Using (\ref{a.26}), (\ref{a.24}) and
(\ref{a.23}) in (\ref{a.9}) we get:
\begin{eqnarray}\label{a.29}
&& \frac x {{\cal F}_2} = z \left(D_3 \frac {{\cal F}_3} {{\cal F}_2} + {\cal
F}_4 - C_3 y\right) - z_0 {\cal F}_4 + \frac 1 {2KR} \nonumber \\
&& + \frac 1 2 \left[\left(x - x_0\right)^2 + y^2 + \left(C_1 x_0 + C_3
z_0\right)^2 + \right. \nonumber \\
&& \left.\left(z - z_0\right)^2\right] + D_4 \frac {{\cal F}_3} {{\cal F}_2} +
x_0 {\cal F}_5.
\end{eqnarray}
We differentiate this by $y$ alone and obtain
\begin{equation}\label{a.30}
x \left(\frac 1 {{\cal F}_2}\right),_y = z \left[D_3 \left(\frac {{\cal F}_3}
{{\cal F}_2}\right),_y - C_3\right] + y + D_4 \left(\frac {{\cal F}_3} {{\cal
F}_2}\right),_y.
\end{equation}
Integrating this back and using (\ref{a.28}) for ${\cal F}_3$ we obtain
\begin{equation}\label{a.31}
\frac 1 {{\cal F}_2} = \frac {\frac 1 2 D_1 y^2 + \left(C_1 D_3 - C_3 D_1\right)
y z + C_1 D_4 y} {D_1 x + D_3 z + D_4} + {\cal F}_6(x,z).
\end{equation}
Using (\ref{a.28}) and (\ref{a.31}) in (\ref{a.29}) we obtain
\begin{eqnarray}\label{a.32}
&& x {\cal F}_6(x, z) = \left(z - z_0\right) {\cal F}_4(x, z) + \frac 1 {2KR} +
x_0 {\cal F}_5(x, z) \nonumber \\
&& + \frac 1 2 \left[\left(x - x_0\right)^2 + \left(C_1 x_0 + C_3 z_0\right)^2
+ \left(z - z_0\right)^2\right] \nonumber \\
&& + \frac 1 {D_1} \left(D_3 z + D_4\right) \left({\cal F}_5(x,z) - {\cal
F}_6(x,z)\right).
\end{eqnarray}
Taking the second derivative of this by $x$ and $t$ we obtain
\begin{equation}\label{a.33}
- \frac {z_{0,t}} {x_{0,t}} {\cal F}_{4,x} - 1 + {\cal F}_{5,x} = 0.
\end{equation}
In Case 1.1.1.1, where $\left(z_{0,t} / x_{0,t}\right),_t \neq 0$, this implies
\begin{equation}\label{a.34}
{\cal F}_4 = {\cal F}_4(z), \qquad {\cal F}_5 = x + {\cal F}_7(z).
\end{equation}
Putting this in (\ref{a.32}) and taking its derivative by $x$ alone, then
integrating back, we get
\begin{equation}\label{a.35}
\left(x + \frac {D_3 z + D_4} {D_1}\right) {\cal F}_6(x, z) = \frac 1 2 x^2 +
\frac {x \left(D_3 z + D_4\right)} {D_1} + {\cal F}_8(z).
\end{equation}
Using this in (\ref{a.32}), then taking the second derivative by $z$ and $t$ we
get
\begin{equation}\label{a.36}
- \frac {z_{0,t}} {x_{0,t}} \left({\cal F}_{4,z} + 1\right) + {\cal F}_{7,z} =
0.
\end{equation}
In the present case this means
\begin{equation}\label{a.37}
{\cal F}_7 = C_4 = {\rm constant}, \qquad {\cal F}_4 = - z + C_5.
\end{equation}
Feeding this information in (\ref{a.32}) and using (\ref{a.24}) we get
\begin{eqnarray}
&& {\cal F}_8(z) - \frac {D_3 C_4} {D_1} z + \frac 1 2 z^2 - C_5 z - \frac {D_4}
{D_1} C_4 = E_1,\ \ \ \ \ \ \  \label{a.38} \\
&& \frac 1 R \left(D_1 x_0 + D_3 z_0 + D_4\right) + C_4 x_0 - C_5 z_0 \nonumber
\\
&& + \frac 1 2 \left[{x_0}^2 + \left(C_1 x_0 + C_3 z_0\right)^2 + {z_0}^2\right]
= E_1, \label{a.39}
\end{eqnarray}
the two expressions being constant because they are equal, while the first one
depends only on $z$ and the second one only on $t$. Equations (\ref{a.38}) --
(\ref{a.39}) define $R(t)$ and ${\cal F}_8(z)$, and are the final solution of
(\ref{a.7}).

Picking up the pieces, we obtain the following formula:
\begin{eqnarray}
{\cal F}_1 &=& \frac {{\cal U}_1}  {{\cal U}_2},  \label{a.40} \\
{\cal U}_1 &\df& \frac 1 2 D_3 \left(x^2 - y^2 - z^2\right) + \left(C_1 D_3 -
C_3 D_1\right) x y \nonumber  \\
&& - D_1 xz + \left(C_4 D_3 + C_5 D_1\right) x - C_3 D_4 y - D_4 z \nonumber \\
&& + C_5 D_4 - D_3 E_1, \label{a.41} \\
{\cal U}_2 &\df& \frac 1 2 D_1 \left(x^2 + y^2 - z^2\right) + D_3 xz
\nonumber \\
&& + \left(C_1 D_3 - C_3 D_1\right) y z + D_4 x + C_1 D_4 y \nonumber \\
&& + \left(C_4 D_3 + C_5 D_1\right) z + C_4 D_4 + D_1 E_1. \label{a.42}
\end{eqnarray}

Now we will deal with (\ref{a.8}), still within Case 1.1.1.1.2.1.

\medskip

{\underline {\bf Case 1.1.1.1.2.1.1: ${\cal G}_2 \neq 0$}}

\medskip

Then (\ref{a.8}) is written as
\begin{eqnarray}\label{a.43}
&& \frac {y - C_1 x_0 - C_3 z_0} {{\cal G}_2} = \left(z - z_0\right) \frac
{{\cal G}_1} {{\cal G}_2} + \frac 1 {2KR} + \frac 1 {2K} \frac {{\cal G}_3}
{{\cal G}_2} \nonumber \\
&& + \frac 1 2 \left[\left(x - x_0\right)^2 + \left(y - C_1 x_0 - C_3
z_0\right)^2 + \left(z - z_0\right)^2\right], \nonumber \\
\end{eqnarray}
We follow exactly the same sequence of steps that we did in solving (\ref{a.9}).
We take the second derivative of (\ref{a.43}) by $y$ and $t$ and obtain, using
(\ref{a.24}):
\begin{eqnarray}\label{a.44}
&& \left(C_1 + C_3 \frac {z_{0,t}} {x_{0,t}}\right) \left(\frac 1 {{\cal
G}_2}\right),_y = \frac {z_{0,t}} {x_{0,t}} \left(\frac {{\cal G}_1} {{\cal
G}_2}\right),_y \\
&& + \left(C_1 + C_3 \frac {z_{0,t}} {x_{0,t}}\right) - \left(D_1 + D_3 \frac
{z_{0,t}} {x_{0,t}}\right) \left(\frac {{\cal G}_3} {{\cal G}_2}\right),_y.
\nonumber
\end{eqnarray}
Since we are in Case 1.1.1.1, the coefficients of $\left(z_{0,t} /
x_{0,t}\right)$ and the remaining terms must balance separately. Moreover, we
are in Case 1.1.1.1.2.1, where $D_1 \neq 0$. Integrating the two equations with
respect to $y$ we obtain
\begin{eqnarray}
&& \frac {{\cal G}_1} {{\cal G}_2} = \left(\frac {C_1 D_3} {D_1} - C_3\right)
\left(y - \frac 1 {{\cal G}_2}\right) +  {\cal G}_4(x,z), \nonumber \\
&& \ \ \ \ \ \ \ \ + D_3 {\cal G}_5(x, z), \label{a.45} \\
&& \frac {{\cal G}_3} {{\cal G}_2} = \frac {C_1} {D_1} \left(y - \frac 1 {{\cal
G}_2}\right) + {\cal G}_5(x,z). \label{a.46}
\end{eqnarray}
We substitute (\ref{a.45}) -- (\ref{a.46}) in (\ref{a.43}) differentiated by $y$
alone and obtain
\begin{eqnarray}\label{a.47}
&& \frac 1 {{\cal G}_2} + y \left(\frac 1 {{\cal G}_2}\right),_y = z \left(\frac
{C_1 D_3} {D_1} - C_3\right) \left[1 - \left(\frac 1 {{\cal
G}_2}\right),_y\right] \nonumber \\
&& \ \ \ \ \ \ \ \ + \frac {C_1 D_4} {D_1} \left[1 - \left(\frac 1 {{\cal
G}_2}\right),_y\right] + y.
\end{eqnarray}
{}From here:
\begin{eqnarray}
&& \frac 1 {{\cal G}_2} = \label{a.48} \\
&& \frac {\frac 1 2 D_1 y^2 + \left(C_1 D_3 - C_3 D_1\right) yz + C_1 D_4 y +
D_1 {\cal G}_6(x,z)} {D_1 y + \left(C_1 D_3 - C_3 D_1\right) z + C_1 D_4}.
\nonumber
\end{eqnarray}
After substituting (\ref{a.45}), (\ref{a.46}), (\ref{a.48}), (\ref{a.23}) and
(\ref{a.24}) in (\ref{a.43}) we obtain
\begin{eqnarray}\label{a.49}
&& \left(z - z_0\right) {\cal G}_4(x, z) + \left(D_1 x_0 + D_3 z_0 + D_4\right)
\frac 1 R \nonumber \\
&& + \frac 1 2 \left[\left(x - x_0\right)^2 + \left(C_1 x_0 + C_3 z_0\right)^2 +
\left(z - z_0\right)^2\right] \nonumber \\
&& + \left(D_1 x_0 +D_3 z + D_4\right){\cal G}_5(x, z) - {\cal G}_6(x,z) = 0.\ \
\ \ \ \ \ \ \ \
\end{eqnarray}
Following further exactly the same scheme that we presented in solving
(\ref{a.7}) we arrive at an equation determining $R(t)$ that is identical with
(\ref{a.39}), i.e. does not put any additional limitation on the $R$ determined
by (\ref{a.39}). Thus, (\ref{a.39}) gives the final condition for the existence
of RLPs in Case 1.1.1.1.2.1.

Putting together all the partial results we get
\begin{eqnarray}
&& {\cal G}_1 = \frac {{\cal U}_3} {{\cal U}_2}, \label{a.50} \\
&& {{\cal U}_3} \df \frac 1 2 \left(C_1 D_3 - C_3 D_1\right) \left(- x^2 + y^2 -
z^2\right) \nonumber \\
&&\ \ \ \  + D_3 xy - D_1 yz + C_3 D_4 x + \left(C_4 D_3 + C_5 D_1\right) y
\nonumber \\
&&\ \ \ \  - C_1 D_4 z + D_4 \left(C_1 C_5 + C_3 C_4\right) \nonumber \\
&&\ \ \ \ - E_1 \left(C_1 D_3 - C_3 D_1\right), \label{a.51}
\end{eqnarray}
where ${\cal U}_2$ is given in (\ref{a.42}). Now the equations
\begin{equation}\label{a.52}
\dr x z = {\cal F}_1(x,y,z), \qquad \dr y z = {\cal G}_1(x,y,z)
\end{equation}
do determine the RLPs, with ${\cal F}_1(x,y,z)$ and ${\cal G}_1(x,y,z)$ given by
(\ref{a.40}) -- (\ref{a.42}) and (\ref{a.50}) -- (\ref{a.51}).\footnote{The
correctness of (\ref{a.39}) -- (\ref{a.42}) and (\ref{a.50}) -- (\ref{a.51}) was
verified by the computer algebra program Ortocartan \cite{Kras2001, KrPe2000}.}

By comparing (\ref{a.24}) and (\ref{a.39}) with (\ref{7.15}) -- (\ref{7.16}) we
see that the latter is contained in (\ref{a.39}) as the subcase $x_0 = 0 = C_3$,
with $D_3, D_4, C_5$ and $E_1$ expressed in terms of $C, A_1, A_2, B_1$ and
$B_2$.

Likewise, comparing (\ref{a.24}) and (\ref{a.39}) with (\ref{9.19}) --
(\ref{9.20}) we see that the latter is contained in (\ref{a.39}) as the subcase
$x_0 = 0 = C_3 = C_5$, with $y_0$ transformed to $z_0$ by the coordinate
transformation $(y, z) = (z', y')$. The other constants in (\ref{a.39}) are
expressed in terms of those from (\ref{9.19}) -- (\ref{9.20}).

These two special Stephani solutions will be contained also in several other
subcases of \{(\ref{a.24}), (\ref{a.39})\} that will appear in the following
text. The spherically symmetric Stephani solution will appear in several places
as well, but each time it will be easy to recognise.

For completeness, we will still consider:

\medskip

{\underline {\bf Case 1.1.1.1.2.1.2: ${\cal G}_2 = 0$}}

\medskip

With $D_1 \neq 0 \neq x_{0,t}$ and $\left(z_{0,t} / x_{0,t}\right),_t \neq 0$
still applying, ${\cal G}_2 = 0$ quickly leads to a contradiction; it is enough
to divide (\ref{a.8}) by $2K$, then differentiate the result by $y$ and $t$ and
use the conclusion in the derivative by $y$ alone. This case is thus empty.

Case 1.1.1.1.2.1 is thereby exhausted, and we go back to (\ref{a.28}) to
consider:

\medskip

{\underline {\bf Case 1.1.1.1.2.2: $D_1 = 0$}}

\medskip

By applying the same scheme as with $D_1 \neq 0$ we obtain formulae for ${\cal
F}_1$, ${\cal G}_1$ and $R$ that are the subcases $D_1 = 0$ of (\ref{a.39}),
(\ref{a.40}) -- (\ref{a.42}) and (\ref{a.50}) -- (\ref{a.51}).

When $C_1 = 0$, we obtain $x_{0,t} = 0$, which contradicts the definition of
Case 1.1.1.

This exhausts Case 1.1.1.1, so we go back to (\ref{a.12}) to consider

\medskip

{\bf \underline {Case 1.1.1.2: $\left(z_{0,t} / x_{0,t}\right),_t = 0$}}

\medskip

Then (\ref{a.12}) becomes
\begin{equation}\label{a.53}
\left(\frac {y_{0,t}} {x_{0,t}}\right),_t - \left[\frac 1 {x_{0,t}} \left(\frac
1 {2K}\right),_t\right],_t \left(\frac {{\cal F}_3} {{\cal F}_2}\right),_y = 0.
\end{equation}

\medskip

{\bf \underline {Case 1.1.1.2.1: $\left[(1 / x_{0,t}) (1 / 2K),_t\right],_t \neq
0$}}

\medskip

For the sake of comparison with the previously considered cases, from this place
up to (\ref{a.65}) we replace the capital letters denoting constants by their
corresponding l.c. letters. Then the solution of (\ref{a.53}) is
\begin{equation}\label{a.54}
\left.\left(\frac {y_{0,t}} {x_{0,t}}\right),_t\right/ \left[\frac 1 {x_{0,t}}
\left(\frac 1 {2K}\right),_t\right],_t = \left(\frac {{\cal F}_3} {{\cal
F}_2}\right),_y = c_1.
\end{equation}

\medskip

{\bf \underline {Case 1.1.1.2.1.1: $c_1 \neq 0$}}

\medskip

We then get
\begin{eqnarray}
&& \frac {c_1} {2K} = d_1 x_0 + y_0 + d_4, \label{a.55} \\
&& {\cal F}_3 = \left(c_1 y + {\cal F}_4(x, z)\right) {\cal F}_2. \label{a.56}
\\
&& z_0 = c_3 x_0, \label{a.57}
\end{eqnarray}
the last one from the definition of Case 1.1.1.2.

We substitute (\ref{a.55}) -- (\ref{a.57}) in (\ref{a.11}). In Case 1.1.1, where
$x_{0,t} \neq 0$, the resulting equation integrates as
\begin{equation}\label{a.58}
c_3 {\cal F}_1 = 1 + \left(d_1 y - {\cal F}_5(x,z)\right) {\cal F}_2.
\end{equation}

\medskip

{\bf \underline {Case 1.1.1.2.1.1.1: $c_3 \neq 0$}}

\medskip

We then use (\ref{a.55}) -- (\ref{a.58}) in (\ref{a.9}) differentiated by $y$
and obtain
\begin{equation}\label{a.59}
\frac 1 {{\cal F}_2} = \frac {\frac 1 2 y^2 + \frac {d_1} {c_3} y z + d_4 y +
{\cal F}_6(x,z)} {x - z / c_3}.
\end{equation}

After substituting (\ref{a.55}) -- (\ref{a.59}) in (\ref{a.9}) we obtain
\begin{eqnarray}\label{a.60}
&& - \frac z {c_3} {\cal F}_5(x, z) + \frac 1 {2KR} + x_0 {\cal F}_5(x,z) -
{\cal F}_6(x,z) \nonumber \\
&& + \frac 1 2 \left[\left(x - x_0\right)^2 + {y_0}^2 + \left(z -
c_3 x_0\right)^2\right] \nonumber \\
&& + \frac 1 {c_1} \left(d_1 x_0 + y_0 + d_4\right) {\cal F}_4(x,z) = 0.
\end{eqnarray}
Following the same sequence of steps that we described below (\ref{a.8}), we
arrive at:
\begin{eqnarray}\label{a.61}
&& \frac 1 {c_1 R} \left(d_1 x_0 + y_0 + d_4\right) + \frac 1 2 \left[\left(1 +
{c_3}^2\right){x_0}^2 + {y_0}^2\right] \nonumber \\
&& + c_5 x_0 + c_6 y_0 = E_1 = {\rm constant}.
\end{eqnarray}
This is equivalent to the subcase $C_3 = 0$ of (\ref{a.39}) transformed by $(y,
z) = (z', y')$ (this transformation interchanges the names of $y_0$ and $z_0$).
Note from (\ref{a.5}) that the interchange of $y$ and $z$ implies the
interchange of $G_y$ and $G_z$, and consequently the transformation $({\cal
F}_1, {\cal G}_1) \to ({\cal F}_1 / {\cal G}_1, 1 / {\cal G}_1)$. Thus the
${\cal F}_1$ and ${\cal G}_1$ for (\ref{a.61}) are obtained from (\ref{a.40}) --
(\ref{a.42}) and (\ref{a.50}) -- (\ref{a.51}) in this way, with the
substitutions $C_3 = 0$ and $(y, z) \to (z, y)$. To have a consistent naming of
the constants, one should do the following replacement in (\ref{a.61}):
\begin{equation}\label{a.62}
\left(d_1 / c_1, 1 / c_1, d_4 / c_1, c_3, c_5, c_6\right) = \left(D_1, D_3, D_4,
C_1, C_4, - C_5\right).
\end{equation}

This is the end of Case 1.1.1.2.1.1.1. We go back to (\ref{a.58}) and consider
now

\medskip

{\bf \underline {Case 1.1.1.2.1.1.2: $c_3 = 0$}}

\medskip

Even though the limit $c_3 = 0$ is singular in (\ref{a.59})-- (\ref{a.60}), by
going through the usual procedure we end up with the subcase $c_3 = 0$ of
(\ref{a.61}). This is also a regular subcase of the current ${\cal F}_1$ and
${\cal G}_1$, calculated in the way explained above.

Thus we go back to (\ref{a.54}) and consider now

\medskip

{\bf \underline {Case 1.1.1.2.1.2: $c_1 = 0$}}

\medskip

Then (\ref{a.55}) does not exist because (\ref{a.54}) only determines $(y_{0,t}
/ x_{0,t}),_t = 0$ and $({\cal F}_3 / {\cal F}_2),_y = 0$. From the first of
these we have
\begin{equation}\label{a.63}
y_0 = C_2 x_0,
\end{equation}
and (\ref{a.56}) still holds with $c_1 = 0$.

\medskip

{\bf \underline {Case 1.1.1.2.1.2.1: $c_3 \neq 0$}}

\medskip

Proceeding as before by consecutive differentiations we end up with the
following equation:
\begin{equation}\label{a.64}
\frac 1 {2KR} + \frac {c_4} {2K} + \frac 1 2 \left(1 + {C_2}^2 + {c_3}^2\right)
{x_0}^2 + c_5 x_0 + c_6 = 0.
\end{equation}
The function $K(t)$ is still arbitrary at this point. Since in the
acceleration-free limit $x_0$ and $KR = k/4$ are constant, we see that $k/R =$
constant in this limit. So, (\ref{a.64}) can reproduce the $k = 0$ FLRW limit.

Running this solution through (\ref{a.8}) with ${\cal G}_2 \neq 0$, with the
current values of $y_0$ and $z_0$, and with $\left[(1 / x_{0,t}) (1 /
2K),_t\right],_t \neq 0$ as appropriate for the current Case 1.1.1.2.1, we
obtain
\begin{equation}\label{a.65}
\frac 1 {2KR} + \frac {E_1} {2K} + \frac 1 2 \left(1 + {C_2}^2 + {c_3}^2\right)
{x_0}^2 + E_2 x_0 + E_3 = 0,
\end{equation}
where $E_i$, $i = 1, 2, 3$, are new arbitrary constants. Choosing $(E_1, E_2,
E_3) = (c_4, c_5, c_6)$ we make (\ref{a.65}) identical to (\ref{a.61}). Then
(\ref{a.64}) does define a subcase of the Stephani model, different from
(\ref{a.39}), that also has RLPs. The corresponding ${\cal F}_1$ and ${\cal
G}_1$ are found to be
\begin{equation}
{\cal F}_1 = \frac {\frac 1 2 \left(x^2 - y^2 - z^2\right) + C_2 x y + C_3 xz +
C_5 x + C_6} {\frac 1 2 C_3 \left(- x^2 - y^2 + z^2\right) + xz + C_2 y z + C_5
z + C_3 C_6}, \label{a.66}
\end{equation}
\begin{equation}
{\cal G}_1 = \frac {\frac 1 2 C_2 \left(x^2 - y^2 + z^2\right) - xy - C_3 yz -
C_5 y - C_2 C_6} {\frac 1 2 C_3 \left(x^2 + y^2 - z^2\right) - xz - C_2 y z -
C_5 z - C_3 C_6}, \label{a.67}
\end{equation}

Even though the subcase $C_3 = 0$ requires separate treatment at intermediate
stages of the calculation, the final formulae for $1 / R$, ${\cal F}_1$ and
${\cal G}_1$ turn out to be contained in (\ref{a.64}), (\ref{a.66}) and
(\ref{a.67}) as the regular subcase $C_3 = 0$. While considering $C_3 = 0$, the
subcase $C_2 = 0$ also requires separate treatment, but in the end leads to $R$,
${\cal F}_1$ and ${\cal G}_1$ given by (\ref{a.64}) -- (\ref{a.67}) with $C_3 =
C_2 = 0$. When $C_2 = 0$, the ${\cal G}_2$ in (\ref{a.8}) must be
zero.\footnote{The correctness of (\ref{a.64}) -- (\ref{a.67}) was verified by
the computer algebra program Ortocartan \cite{Kras2001, KrPe2000}.}

When we run the general expression for (\ref{a.64}) through (\ref{a.8}) with
${\cal G}_2 = 0$, we quickly obtain $c_3 = C_2 = 0$ as a necessary consequence.

Having thus exhausted Case 1.1.1.2.1 we go back to (\ref{a.53}) and consider

\medskip

{\bf \underline {Case 1.1.1.2.2: $\left[(1 / x_{0,t}) (1 / 2K),_t\right],_t =
0$}}

\medskip

Then, using (\ref{a.53}) and the definitions of Cases 1.1.1.2 and 1.1.1.2.2 we
obtain:
\begin{equation}\label{a.68}
\frac 1 {2K} = D_1 x_0 + D_2, \qquad y_0 = C_2 x_0, \qquad z_0 = C_3 x_0.
\end{equation}
Proceeding further by the ordinary scheme we obtain from (\ref{a.9}):
\begin{equation}\label{a.69}
\frac 1 R \left(D_1 x_0 + D_2\right) + \frac 1 2 \left(1 + {C_2}^2 +
{C_3}^2\right) {x_0}^2 + C_4 x_0 = C_5.
\end{equation}
This is formally a subcase of (\ref{a.39}), but it has $G_x = G_y = G_z = 0$,
and so belongs in Case 2 considered further on.

This exhausts Case 1.1.1, so we go back to (\ref{a.11}) and consider

\medskip

{\bf \underline {Case 1.1.2: $x_{0,t} = 0$}}

\medskip

Then (\ref{a.11}) becomes
\begin{equation}\label{a.70}
z_{0,t} \left(\frac {{\cal F}_1} {{\cal F}_2}\right),_y + y_{0,t} - \left(\frac
1 {2K}\right),_t \left(\frac {{\cal F}_3} {{\cal F}_2}\right),_y = 0.
\end{equation}

\medskip

{\bf \underline {Case 1.1.2.1: $y_{0,t} \neq 0$}}

\medskip

Then we divide (\ref{a.70}) by $y_{0,t}$ and differentiate the result by $t$,
obtaining
\begin{equation}\label{a.71}
\left(\frac {z_{0,t}} {y_{0,t}}\right),_t \left(\frac {{\cal F}_1} {{\cal
F}_2}\right),_y - \left[\frac 1 {y_{0,t}} \left(\frac 1 {2K}\right),_t\right],_t
\left(\frac {{\cal F}_3} {{\cal F}_2}\right),_y = 0.
\end{equation}

\medskip

{\bf \underline {Case 1.1.2.1.1: $\left(z_{0,t} / y_{0,t}\right),_t \neq 0$}}

\medskip

Then we divide (\ref{a.71}) by $\left(z_{0,t} / y_{0,t}\right),_t$ and
differentiate the result by $t$.

\medskip

{\bf \underline {Case 1.1.2.1.1.1: $\chi_1(t) \neq 0$}}

\medskip

where
\begin{equation}\label{a.72}
\chi_1(t) \df \left\{\left[\frac 1 {y_{0,t}} \left(\frac 1
{2K}\right),_t\right],_t \right/\left.\left(\frac {z_{0,t}}
{y_{0,t}}\right),_t\right\},_t.
\end{equation}
Then $\left({\cal F}_3 / {\cal F}_2\right),_y = 0$ and (\ref{a.71}) immediately
implies $\left({\cal F}_1 / {\cal F}_2\right),_y = 0$. However, then
(\ref{a.70}) gives a contradiction with $y_{0,t} \neq 0$. This case is thus
empty, so we proceed to consider the complementary

\medskip

{\bf \underline {Case 1.1.2.1.1.2: $\chi_1(t) = 0$}}

\medskip

Then we have
\begin{equation}\label{a.73}
\frac 1 {2K} = D_1 + D_2 y_0 + D_3 z_0,
\end{equation}
and from (\ref{a.71})
\begin{equation}\label{a.74}
\frac {{\cal F}_1} {{\cal F}_2} = D_3 \frac {{\cal F}_3} {{\cal F}_2} + {\cal
F}_4(x, z).
\end{equation}
Proceeding from (\ref{a.9}) by the usual method we arrive at
\begin{eqnarray}\label{a.75}
&& \frac 1 R \left(D_1 + D_2 y_0 + D_3 z_0\right) + \frac 1 2 \left({y_0}^2 +
{z_0}^2\right) \nonumber \\
&& + C_4 y_0 + C_5 z_0 = E_1.
\end{eqnarray}
This is equivalent to the subcase $C_1 = C_3 = 0$ of (\ref{a.39}) under the
coordinate transformation $(x, y) = (y', x')$ that interchanges the names of
$x_0$ and $y_0$. Note, by looking at (\ref{a.7}) -- (\ref{a.8}), that the
interchange of $x$ and $y$ implies the interchange of ${\cal F}_1$ and ${\cal
G}_1$. Thus, the corresponding ${\cal F}_1$ and ${\cal G}_1$ are found from
(\ref{a.40}) -- (\ref{a.42}) and (\ref{a.50}) -- (\ref{a.51}) as, respectively,
the old ${\cal G}_1$ and ${\cal F}_1$ in the limit $C_1 = C_3 = 0$ with $(x, y)
\to (y', x')$, and with the additional renaming $(C_5, D_1, D_4) \to (- C_5,
D_2, D_1)$ (compare (\ref{a.39}) with (\ref{a.75})).

So we go back to (\ref{a.71}) to consider

\medskip

{\bf \underline {Case 1.1.2.1.2: $\left(z_{0,t} / y_{0,t}\right),_t = 0$}}

\medskip

Then
\begin{equation}\label{a.76}
z_0 = C_3 y_0.
\end{equation}

\medskip

{\bf \underline {Case 1.1.2.1.2.1: $\left[\left( 1 / y_{0,t}\right) \left(1 /
2K\right),_t\right],_t \neq 0$}}

\medskip

Then (\ref{a.71}) gives
\begin{equation}\label{a.77}
{\cal F}_3 / {\cal F}_2 = {\cal F}_4(x, z),
\end{equation}
and (\ref{a.70}) implies
\begin{equation}\label{a.78}
C_3 \frac {{\cal F}_1} {{\cal F}_2} = - y + {\cal F}_5(x, z).
\end{equation}
Note that $C_3 \neq 0$, or else (\ref{a.78}) is a contradiction.

Proceeding from (\ref{a.9}) by the usual routine we obtain
\begin{equation}\label{a.79}
\frac 1 {2KR} + \frac {C_4} {2K} + \frac 1 2 \left(1 + {C_3}^2\right) {y_0}^2 +
C_5 y_0 + C_6 = 0,
\end{equation}
with $K(t)$ undetermined. This is equivalent to the subcase $C_2 = 0$ of
(\ref{a.64}). The transformation from (\ref{a.64}) and (\ref{a.66}) --
(\ref{a.67}) to the current case is $(x, y) = (y', x')$, with the accompanying
renaming $(x_0, y_0) \to (y_0, x_0)$. As explained under (\ref{a.75}), the
interchange of $x$ and $y$ implies the interchange of ${\cal F}_1$ and ${\cal
G}_1$, so the current ${\cal F}_1$ and ${\cal G}_1$ are obtained as the ${\cal
G}_1$ and ${\cal F}_1$, respectively, of (\ref{a.66}) -- (\ref{a.67}) with  $(x,
y) \to (y, x)$ and $C_2 = 0$.

So, we go back to (\ref{a.71}) once more and consider the second possibility:

\medskip

{\bf \underline {Case 1.1.2.1.2.2: $\left[\left( 1 / y_{0,t}\right) \left(1 /
2K\right),_t\right],_t = 0$}}

\medskip

Then
\begin{equation}\label{a.80}
\frac 1 {2K} = D_1 + D_2 y_0,
\end{equation}
and from (\ref{a.70}), since $y_{0,t} \neq 0$,
\begin{equation}\label{a.81}
C_3 \frac {{\cal F}_1} {{\cal F}_2} = D_2 \frac {{\cal F}_3} {{\cal F}_2} - y +
{\cal F}_4(x, z).
\end{equation}

\medskip

{\bf \underline {Case 1.1.2.1.2.2.1: $C_3 \neq 0$}}

\medskip

Then, by the usual routine, (\ref{a.81}) used in (\ref{a.9}) leads to
\begin{equation}\label{a.82}
\frac 1 R \left(D_1 + D_2 y_0\right) + \frac 1 2 \left(1 + {C_3}^2\right)
{y_0}^2 + C_4 y_0 = C_5,
\end{equation}
which is formally the subcase $C_5 = D_3 = 0$, $z_0 = C_3 x_0$ of (\ref{a.75}).
However, in this case we have $G_x = G_y = G_z = 0$ in (\ref{11.5}) --
(\ref{11.6}). Thus, it is in fact excluded from the present consideration and
will appear later, when we consider Case 2.

\medskip

{\bf \underline {Case 1.1.2.1.2.2.2: $C_3 = 0$}}

\medskip

Then the subcase $C_3 = 0$ of (\ref{a.82}) is obtained, so it belongs in Case 2,
too.

This exhausts Case 1.1.2.1, so we go back to (\ref{a.70}) to consider

\medskip

{\bf \underline {Case 1.1.2.2: $y_{0,t} = 0$}}

\medskip

Then $y_0 = 0$ by a transformation of $y$, and in (\ref{a.70}) we consider

\medskip

{\bf \underline {Case 1.1.2.2.1: $z_{0,t} \neq 0$}}

\medskip

Then (\ref{a.70}) may be written as
\begin{equation}\label{a.83}
\left(\frac {{\cal F}_1} {{\cal F}_2}\right),_y - \frac 1 {z_{0,t}} \left(\frac
1 {2K}\right),_t \left(\frac {{\cal F}_3} {{\cal F}_2}\right),_y = 0.
\end{equation}
When $\left\{(1 / z_{0,t}) \left[1 / (2K)\right],_t\right\},_t \neq 0$, this
yields $\left({\cal F}_1 / {\cal F}_2\right),_y$ $= \left({\cal F}_3 / {\cal
F}_2\right),_y = 0$, and the usual routine leads from (\ref{a.9}) to
\begin{equation}\label{a.84}
\frac 1 {2KR} + \frac {C_4} {2K} + \frac 1 2 {z_0}^2 + C_5 z_0 + C_6 = 0,
\end{equation}
with $K(t)$ undetermined. This is the subcase $c_3 = C_2 = 0$ of (\ref{a.64})
obtained by the coordinate transformation $(x, z) = (z', x')$ that interchanges
$x_0$ with $z_0$. Note from (\ref{a.5}) that the interchange of $x$ and $z$
implies the interchange of $G_x$ and $G_z$, and consequently the transformation
$({\cal F}_1, {\cal G}_1) \to (1 / {\cal F}_1, {\cal G}_1 / {\cal F}_1)$. Thus,
the ${\cal F}_1$ and ${\cal G}_1$ for (\ref{a.84}) are obtained from
(\ref{a.66}) -- (\ref{a.67}) in this way, with the substitutions $C_2 = C_3 = 0$
and $(x, z) \to (z, x)$.

Thus we assume $\left[\frac 1 {z_{0,t}} \left(\frac 1 {2K}\right),_t\right],_t =
0$ in (\ref{a.83}). Then
\begin{eqnarray}
\frac 1 {2K} &=& D_1 + D_3 z_0, \label{a.85} \\
\frac {{\cal F}_1} {{\cal F}_2} &=& D_3 \frac {{\cal F}_3} {{\cal F}_2} + {\cal
F}_4(x, z). \label{a.86}
\end{eqnarray}
This leads from (\ref{a.9}) to
\begin{equation}\label{a.87}
\frac 1 R \left(D_1 + D_3 z_0\right) + \frac 1 2 {z_0}^2 + C_5 z_0 = E_1.
\end{equation}
This is the subcase $y_0 = 0$ of (\ref{a.75}). As with (\ref{a.82}), in this
case we have $G_x = G_y = G_z = 0$, and so it is excluded from the present
consideration -- it will appear in Case 2.

So we go back to (\ref{a.71}) and consider

\medskip

{\bf \underline {Case 1.1.2.2.2: $z_{0,t} = 0$}}

\medskip

With $x_{0,t} = y_{0,t} = z_{0,t} = 0$ now being the case, we are in the
spherically symmetric subcase of the Stephani solution. Then (\ref{a.9}) may be
written as
\begin{equation}\label{a.88}
\frac x {{\cal F}_2} = z \frac {{\cal F}_1} {{\cal F}_2} + \frac 1 {2KR} + \frac
1 2 \left(x^2 + y^2 + z^2\right) + \frac 1 {2K} \frac {{\cal F}_3} {{\cal F}_2}.
\end{equation}
The $t$-derivative of this is
\begin{equation}\label{a.89}
\left(\frac 1 {2KR}\right),_t + \left(\frac 1 {2K}\right),_t \frac {{\cal F}_3}
{{\cal F}_2} = 0.
\end{equation}
If $(2KR),_t = 0$, then this is simply the FLRW family of models that we need
not investigate. Consequently, we take $(2KR),_t \neq 0$. Then both the other
factors must be nonzero, and
\begin{equation}\label{a.90}
\left.\left(\frac 1 {2KR}\right),_t\right/ \left(\frac 1 {2K}\right),_t = -
\frac {{\cal F}_3} {{\cal F}_2} = C_1.
\end{equation}
This is the subcase of the spherically symmetric Stephani solution that we
identified in Sec. \ref{spherlps}, and the equation above is consistent with
(\ref{5.10}). All of its null geodesics are RLPs. The FLRW limit of this model
is $C_1 = 0$, and it includes the whole FLRW family.

Equation (\ref{a.90}) is the final solution of (\ref{a.7}). To prevent
(\ref{a.8}) from imposing any limitations on it, it is sufficient to choose
${\cal G}_2 = {\cal G}_3 = 0$, ${\cal G}_1 = y/z$.

This exhausts Case 1.1, so we go back to (\ref{a.7}) and take

\medskip

{\bf \underline {Case 1.2: ${\cal F}_2 = 0$}}

\medskip

Then (\ref{a.7}) can be rewritten as
\begin{equation}\label{a.91}
x - x_0 = \left(z - z_0\right) {\cal F}_1 + \frac 1 {2K} {\cal F}_3.
\end{equation}
We differentiate this by $y$ and $t$ and obtain
\begin{equation}\label{a.92}
-z_{0,t} {\cal F}_{1,y} + \left(\frac 1 {2K}\right),_t {\cal F}_{3,y} = 0.
\end{equation}

\medskip

{\bf \underline {Case 1.2.1: $z_{0,t} \neq 0$}}

\medskip

Then from (\ref{a.91})
\begin{equation}\label{a.93}
- {\cal F}_{1,y} + \frac 1 {z_{0,t}} \left(\frac 1 {2K}\right),_t {\cal F}_{3,y}
= 0.
\end{equation}
When $\left[\frac 1 {z_{0,t}} \left(\frac 1 {2K}\right),_t\right],_t \neq 0$, we
have ${\cal F}_{1,y} = {\cal F}_{3,y} = 0$, and then we differentiate
(\ref{a.91}) by $x$ to get
\begin{equation}\label{a.94}
1 = \left(z - z_0\right) {\cal F}_{1,x} + \frac 1 {2K} {\cal F}_{3,x}.
\end{equation}
Differentiating this by $t$ we get ${\cal F}_{1,x} = {\cal F}_{3,x} = 0$ in
consequence of the assumptions about the functions of $t$, but this is a
contradiction with (\ref{a.94}). This means that $\left[\frac 1 {z_{0,t}}
\left(\frac 1 {2K}\right),_t\right],_t = 0$, i.e.
\begin{equation}
\frac 1 {2K} = D_3 z_0 + D_4, \label{a.95} \\
\end{equation}
Continuing from (\ref{a.91}) by the usual procedure we get the following final
solution:
\begin{eqnarray}
x_0 &=& C_2 z_0, \label{a.96} \\
{\cal F}_1 &=& \frac {D_3 x + C_2 D_4} {D_4 + D_3 z}, \label{a.97} \\
{\cal F}_3 &=& \frac {x - C_2 z} {D_4 + D_3 z}, \label{a.98}
\end{eqnarray}
and we next have to verify (\ref{a.8}). We first try

\medskip

{\bf \underline {Case 1.2.1.1: ${\cal G}_2 \neq 0$}}

\medskip

Then (\ref{a.8}) can be written as
\begin{eqnarray}\label{a.99}
&& \frac {y - y_0} {{\cal G}_2} = \left(z - z_0\right) \frac {{\cal G}_1} {{\cal
G}_2} + \frac 1 {2KR} \\
&& + \frac 1 2 \left[\left(x - C_2 z_0\right)^2 + \left(y - y_0\right)^2 +
\left(z - z_0\right)^2\right] + \frac 1 {2K} \frac {{\cal G}_3} {{\cal G}_2},
\nonumber
\end{eqnarray}
Using (\ref{a.95}) and (\ref{a.96}) we take the second derivative of this by $y$
and $t$. Since we are in Case 1.2.1, where $z_{0,t} \neq 0$, we get
\begin{equation}\label{a.100}
\left(\frac {{\cal G}_1} {{\cal G}_2}\right),_y - D_3 \left(\frac {{\cal G}_3}
{{\cal G}_2}\right),_y = \frac {y_{0,t}} {z_{0,t}} \left[\left(\frac 1 {{\cal
G}_2}\right),_y - 1\right].
\end{equation}

\medskip

{\bf \underline {Case 1.2.1.1.1: $\left(y_{0,t}/z_{0,t}\right),_t \neq 0$}}

\medskip

Then (\ref{a.100}) solves as
\begin{eqnarray}
\frac 1 {{\cal G}_2} &=& y + {\cal G}_4(x,z), \label{a.101} \\
{\cal G}_1 &=& D_3 {\cal G}_3 + \frac {{\cal G}_5(x, z)} {y + {\cal G}_4(x, z)}.
\label{a.102}
\end{eqnarray}
Proceeding from (\ref{a.99}) by the usual routine we find ${\cal G}_3$ and the
arbitrary functions of $(x, z)$, and we end up with
\begin{eqnarray}\label{a.103}
&& \frac 1 R \left(D_4 + D_3 z_0\right) + \frac 1 2 \left[{y_0}^2 + \left(1 +
{C_2}^2\right) {z_0}^2\right] \nonumber \\
&& \ \ \ \ \ \ + C_4 y_0 + C_5 z_0 = E_1.
\end{eqnarray}
This is equivalent to the subcase $D_1 = C_1 = 0$ of (\ref{a.39}) under the
coordinate transformation $(x, y) = (y', x')$ that interchanges $x_0$ with
$y_0$. The remark under (\ref{a.75}) about the transformation of ${\cal F}_1$
and ${\cal G}_1$ applies also here.

We go back to (\ref{a.100}) and consider

\medskip

{\bf \underline {Case 1.2.1.1.2: $\left(y_{0,t}/z_{0,t}\right),_t = 0$}}

\medskip

Then
\begin{eqnarray}
y_0 &=& C_3 z_0, \label{a.104} \\
\frac {{\cal G}_1} {{\cal G}_2} &=& D_3 \frac {{\cal G}_3} {{\cal G}_2} + C_3
\left(\frac 1 {{\cal G}_2} - y\right) + {\cal G}_4(x, z). \label{a.105}
\end{eqnarray}
In the usual way this leads from (\ref{a.99}) to
\begin{equation}\label{a.106}
\frac 1 R \left(D_4 + D_3 z_0\right) + \frac 1 2 \left(1 + {C_2}^2 +
{C_3}^2\right) {z_0}^2 - C_5 z_0 = E_1,
\end{equation}
which is formally the subcase $C_1 = C_4 = D_1 = 0, x_0 = C_2 z_0$ of
(\ref{a.39}), but belongs to the class with $G_x = G_y = G_z = 0$ considered
under Case 2.

This exhausts Case 1.2.1.1., so we consider

\medskip

{\bf \underline {Case 1.2.1.2: ${\cal G}_2 = 0$}}

\medskip

Then (\ref{a.8}) becomes, using (\ref{a.95})
\begin{equation}\label{a.107}
y - y_0 = \left(z - z_0\right) {\cal G}_1 + \left(D_4 + D_3 z_0\right) {\cal
G}_3.
\end{equation}
By the normal routine this leads to
\begin{eqnarray}
y_0 &=& C_3 z_0, \label{a.108} \\
{\cal G}_1 &=& \frac {D_3 y + C_3 D_4} {D_4 + D_3 z}, \label{a.109} \\
{\cal G}_3 &=& \frac {y - C_3 z} {D_4 + D_3 z}, \label{a.110}
\end{eqnarray}
which does not impose any limitation on (\ref{a.95}) -- (\ref{a.98}) and leaves
$R$ undetermined. Thus, this subcase has a nontrivial FLRW limit. We have
\begin{eqnarray}
V &=& \frac 1 R + \frac k {4R} \left[x^2 + y^2 + z^2 - 2 \left(C_2 x + C_3 y +
z\right) z_0\right. \nonumber \\
&&\ \ \ \ \ \  + \left.\left(1 + {C_2}^2 + {C_3}^2\right) {z_0}^2\right],
\label{a.111} \\
k &=& \frac {2 R} {D_4 + D_3 z_0}. \label{a.112}
\end{eqnarray}
The flat FLRW model is contained here in the limit $(D_4, D_2) \to \infty$. In
the general case, an orthogonal transformation of $(x, y, z)$ may be used to
achieve $C_2 = C_3 = 0$ ($z_0$ is then transformed to $\widetilde{z}_0 = \sqrt{1
+ {C_2}^2 + {C_3}^2} z_0$), and then the model is seen to be axially symmetric,
with the orbits of symmetry in the new $(x, y)$ plane. Assuming $C_2 = C_3 = 0$
and using (\ref{a.97}), (\ref{a.109}) and (\ref{a.111}) in (\ref{a.1}) we get
\begin{equation}\label{a.113}
\dr x z = \frac {D_3 x} {D_4 + D_3 z}, \qquad \dr y z = \frac {D_3 y} {D_4 + D_3
z}
\end{equation}
as the equations defining the RLPs in this case. As is easy to see, they obey $y
\dril x z - x \dril y z = 0$, and so are, in the $(x,y)$ surface, straight lines
passing through the symmetry axis $x = y = 0$.

Thereby, Case 1.2.1 is exhausted, so we go back to (\ref{a.92}) to consider

\medskip

{\bf \underline {Case 1.2.2: $z_{0,t} = 0$}}

\medskip

This is equivalent to $z_0 = 0$. Then (\ref{a.92}) leads to two further
subcases:

\medskip

{\bf \underline {Case 1.2.2.1: $\left(1 / 2K\right),_t \neq 0$}}

\medskip

By the usual method we obtain from (\ref{a.92}) and (\ref{a.91}):
\begin{equation}\label{a.114}
{\cal F}_1 = \frac {x - E_4} z, \qquad {\cal F}_3 = E_3, \qquad \frac {E_3} {2K}
= E_4 - x_0,
\end{equation}
with $R$ undetermined. This is the final solution of (\ref{a.7}).

For checking (\ref{a.8}) we have to consider separately

\medskip

{\bf \underline {Case 1.2.2.1.1: $E_3 \neq 0$}}

\medskip

Then (\ref{a.114}) determines $K$, and from the definition of Case 1.2.2.1 it
follows that $x_{0,t} \neq 0$. But to continue, we have to separately consider
${\cal G}_2$ being zero or not.

\medskip

{\bf \underline {Case 1.2.2.1.1.1: ${\cal G}_2 \neq 0$}}

\medskip

Then (\ref{a.8}) is written as
\begin{eqnarray}\label{a.115}
&& \frac {y - y_0} {{\cal G}_2} = z \frac {{\cal G}_1} {{\cal G}_2} + \frac 1
{2KR} \\
&& + \frac 1 2 \left[\left(x - x_0\right)^2 + \left(y - y_0\right)^2 +
z^2\right] + \frac 1 {E_3} \left(E_4 - x_0\right) \frac {{\cal G}_3} {{\cal
G}_2}. \nonumber
\end{eqnarray}
Taking the derivative of this by $y$ and $t$, and knowing that $x_{0,t} \neq 0$,
we get
\begin{equation}\label{a.116}
\frac {y_{0,t}} {x_{0,t}} \left[\left(\frac 1 {{\cal G}_2}\right),_y - 1\right]
= \frac 1 {E_3} \left(\frac {{\cal G}_3} {{\cal G}_2}\right),_y.
\end{equation}

\medskip

{\bf \underline {Case 1.2.2.1.1.1.1: $\left(y_{0,t} / x_{0,t}\right),_t \neq
0$}}

\medskip

Then
\begin{equation}\label{a.117}
\frac 1 {{\cal G}_2} = y + {\cal G}_4(x,z), \qquad \frac {{\cal G}_3} {{\cal
G}_2} = {\cal G}_5(x, z).
\end{equation}
Proceeding from (\ref{a.115}) in the usual way we obtain
\begin{equation}\label{a.118}
\frac 1 {E_3 R} \left(E_4 - x_0\right) + \frac 1 2 \left({x_0}^2 +
{y_0}^2\right) + C_4 x_0 - C_5 y_0 = E_1.
\end{equation}
This is the subcase of (\ref{a.39}) that results when we take $C_1 = C_3 = D_3 =
0$ in (\ref{a.39}) and interchange $y_0$ with $z_0$. How the new ${\cal F}_1$
and ${\cal G}_1$ are calculated after such a transformation is explained under
(\ref{a.61}). To have a consistent naming of the constants, one must take in
(\ref{a.39}) $(D_1, D_4) = (- 1 / E_3, E_4 / E_3)$.

We go back to (\ref{a.116}) and consider

\medskip

{\bf \underline {Case 1.2.2.1.1.1.2: $\left(y_{0,t} / x_{0,t}\right),_t = 0$}}

\medskip

Then from the definition above
\begin{equation}\label{a.119}
y_0 = C_2 x_0,
\end{equation}
and by integrating (\ref{a.116})
\begin{equation}\label{a.120}
\frac {{\cal G}_3} {{\cal G}_2} = C_2 E_3 \left(\frac 1 {{\cal G}_2} - y\right)
+ E_3 {\cal G}_4 (x, z).
\end{equation}
By continuing from (\ref{a.115}) in the usual way we get finally
\begin{equation}\label{a.121}
\frac 1 {E_3 R} \left(E_4 - x_0\right) + \frac 1 2 \left(1 + {C_2}^2\right)
{x_0}^2 + C_4 x_0 = E_1.
\end{equation}
This is formally the subcase $y_0 = C_2 x_0$ of (\ref{a.118}), again with some
renaming of the constants. However, it has $G_x = G_y = G_z = 0$, so in fact it
belongs in Case 2 considered further on.

This completes Case 1.2.2.1.1.1, so we go back to (\ref{a.114}) to consider

\medskip

{\bf \underline {Case 1.2.2.1.1.2: ${\cal G}_2 = 0$}}

\medskip

Then (\ref{a.8}) is written as
\begin{equation}\label{a.122}
y - y_0 = z {\cal G}_1 + \frac 1 {E_3} \left(E_4 - x_0\right) {\cal G}_3.
\end{equation}

\medskip

{\bf \underline {Case 1.2.2.1.1.2.1: $x_{0,t} \neq 0$}}

\medskip

Then the final solution of (\ref{a.122}) is
\begin{equation}\label{a.123}
y_0 = C_2 x_0, \qquad {\cal G}_1 = \frac 1 z \left(y - E_4 C_2\right), \qquad
{\cal G}_3 = C_2 E_3,
\end{equation}
where $C_2$ is allowed to be zero. This imposes no additional conditions on
(\ref{a.114}), and $R$ remains undetermined. The function $V$ is in this case
\begin{eqnarray}\label{a.124}
&& V = \frac 1 R + \frac {E_3} {2 \left(E_4 - x_0\right)} \left[x^2 + y^2 +
z^2\right. \nonumber \\
&& \left.- 2 x_0 \left(x + C_2 y\right) + \left(1 + {C_2}^2\right)
{x_0}^2\right].
\end{eqnarray}
This is equivalent to (\ref{a.111}) -- (\ref{a.112}), which is seen when we
transform $C_2$ to $0$ by a rotation in the $(x, y)$ plane. Subsequently, the
new $(x, z)$ have to be transformed by $(x, z) = (z', x')$. How the new ${\cal
F}_1$ and ${\cal G}_1$ are calculated after such a transformation is explained
under (\ref{a.84}).

We go back to (\ref{a.122}) to consider

\medskip

{\bf \underline {Case 1.2.2.1.1.2.2: $x_{0,t} = 0$}}

\medskip

Then (\ref{a.122}) implies $y_{0,t} = 0$. With $x_{0,t} = y_{0,t} = z_{0,t} = 0$
we are then in the spherically symmetric subcase of the Stephani solution, but
not in its full generality. Since (\ref{a.114}) still applies, with $x_0 =$
constant we have $K \equiv k/(4R) =$ constant, and this corresponds to the
subcase $A_2 = 0$ of (\ref{5.10}). Since this has already been investigated, we
go back to (\ref{a.114}) to consider

\medskip

{\bf \underline {Case 1.2.2.1.2: $E_3 = 0$}}

\medskip

Then (\ref{a.114}) leaves $K$ still arbitrary, but implies $x_0 = E_4$. By a
transformation of $x$ this can be changed to
\begin{equation}\label{a.125}
x_0 = E_4 = 0.
\end{equation}
With this, we go on to consider (\ref{a.8}). Recall that we are still in Case
1.2.2, where also $z_0 = 0$.

\medskip

{\bf \underline {Case 1.2.2.1.2.1: ${\cal G}_2 \neq 0$}}

\medskip

Then (\ref{a.8}) is written as
\begin{eqnarray}\label{a.126}
&& \frac {y - y_0} {{\cal G}_2} = z \frac {{\cal G}_1} {{\cal G}_2} + \frac 1
{2KR} \\
&& + \frac 1 2 \left[x^2 + \left(y - y_0\right)^2 + z^2\right] + \frac 1 {2K}
\frac {{\cal G}_3} {{\cal G}_2}. \nonumber
\end{eqnarray}
Taking the derivative of this by $y$ and $t$ we get
\begin{equation}\label{a.127}
- y_{0,t} \left(\frac 1 {{\cal G}_2}\right),_y = - y_{0,t} + \left(\frac 1
{2K}\right),_t \left(\frac {{\cal G}_3} {{\cal G}_2}\right),_y.
\end{equation}
We are still in Case 1.2.2.1, where $\left(1 / 2K\right),_t \neq 0$, so we
divide (\ref{a.127}) by $\left(1 / 2K\right),_t$ and consider first

\medskip

{\bf \underline {Case 1.2.2.1.2.1.1: $\left[y_{0,t} / \left(1 /
2K\right),_t\right],_t \neq 0$}}

\medskip

Then (\ref{a.127}) implies $\left(1 / {\cal G}_2\right),_y - 1 = \left({\cal
G}_3 / {\cal G}_2\right),_y = 0$. This leads to a simple coordinate transform of
(\ref{a.84}) ($z_0$ replaced by $y_0$), which is a subcase of (\ref{a.64}).

We go back to (\ref{a.127}) and consider

\medskip

{\bf \underline {Case 1.2.2.1.2.1.2: $\left[y_{0,t} / \left(1 /
2K\right),_t\right],_t = 0$}}

\medskip

We have then $y_0 = D_1 / (2K) + D_2$. We do not have to consider the case $D_1
= 0$ because this would mean $y_0 = 0$ in addition to $x_0 = z_0 = 0$, and we
would be back in the spherically symmetric Stephani solution, considered in Sec.
\ref{spherlps}. So the new situation arises only when $D_1 \neq 0$, and then we
can rewrite the last formula as
\begin{equation}\label{a.128}
\frac 1 {2K} = D_1 + D_2 y_0,
\end{equation}
obtaining from (\ref{a.127})
\begin{equation}\label{a.129}
\frac 1 {{\cal G}_2} = y - D_2 \frac {{\cal G}_3} {{\cal G}_2} + {\cal G}_4 (x,
z).
\end{equation}
This leads to the subcase $y_0 = 0$ of (\ref{a.118}), with $x_0$ subsequently
changed to $y_0$. This again has $G_x = G_y = G_z  = 0$ and belongs in Case 2.

Case 1.2.2.1.2.1 is now exhausted, so we go back to (\ref{a.125}) and consider

\medskip

{\bf \underline {Case 1.2.2.1.2.2: ${\cal G}_2 = 0$}}

\medskip

Then (\ref{a.8}) becomes
\begin{equation}\label{a.130}
y - y_0 = z {\cal G}_1 + \frac 1 {2K} {\cal G}_3.
\end{equation}
This leads to
\begin{equation}\label{a.131}
{\cal G}_1 = \frac {y - C_4} z, \qquad {\cal G}_3 = C_5, \qquad \frac {C_5} {2K}
= C_4 - y_0.
\end{equation}
When $C_5 \neq 0$ this is equivalent to (\ref{a.123}) -- (\ref{a.124}), with the
constants in (\ref{a.131}) related to those in (\ref{a.123}) by $(C_4, C_5) =
C_2 (E_4, E_3)$. When $C_5 = 0$, we are back in the spherically symmetric
subclass.

This completes Case 1.2.2.1. We have to go back as far as (\ref{a.92}) and
consider

\medskip

{\bf \underline {Case 1.2.2.2: $\left(1 / 2K\right),_t = 0$}}

\medskip

With $z_{0,t} = 0$ now being considered (Case 1.2.2), (\ref{a.92}) is fulfilled
identically, and (\ref{a.91}) immediately implies $x_{0,t} = 0$. Taking $x_0 =
z_0 = 0$ and $1/2K = B =$ constant in (\ref{a.91}) we get
\begin{equation}\label{a.132}
x = z {\cal F}_1 + B {\cal F}_3,
\end{equation}
with no limitation on $R$, and this is the final solution of (\ref{a.7}).
Running this through (\ref{a.8}) with ${\cal G}_2 \neq 0$ we obtain
\begin{equation}\label{a.133}
\frac B R + \frac 1 2 {y_0}^2 + C_4 y_0 = C_5.
\end{equation}
This case again has $G_x = G_y = G_z = 0$, and so belongs in Case 2.

Running (\ref{a.132}) through (\ref{a.8}) with ${\cal G}_2 = 0$ we immediately
get $y_{0,t} = 0$ and no limitation on $R$. This is a subcase of the spherically
symmetric Stephani solution, the same one that we obtained in Case
1.2.2.1.1.2.2.

This completes Case 1.2, and the whole Case 1. So we go back to (\ref{11.5}) --
(\ref{11.6}) and consider

\medskip

{\bf \underline {Case 2: $G_z = 0$}}

\medskip

Then we immediately have $G_x = G_y = 0$ in (\ref{11.5}) -- (\ref{11.6}). This
means that (\ref{11.5}) and (\ref{11.6}) are fulfilled for any $\dril x z$ and
$\dril y z$, i.e. that all null geodesics are RLPs in this case.

Multiplying the equations $G_x = G_y = G_z = 0$ by ${V,_t}^2$ we turn them into
polynomials in $(x, y, z)$. Taking the coefficients of $x^2$ in each polynomial
we get\footnote{These formulae were calculated using the algebraic computer
program Ortocartan \cite{Kras2001, KrPe2000}.}
\begin{eqnarray}
&& - 2KK,_{t t}x_{0,t} + 2Kx_{0,t t}K,_{t} + 4{K,_{t}}^{2}x_{0,t} = 0, \ \ \ \ \
\ \label{a.134} \\
&& 2KK,_{t t}y_{0,t} - 2Ky_{0,t t}K,_{t} - 4{K,_{t}}^{2}y_{0,t} = 0,
\label{a.135} \\
&& 2KK,_{t t}z_{0,t} - 2Kz_{0,t t}K,_{t} - 4{K,_{t}}^{2}z_{0,t} = 0.
\label{a.136}
\end{eqnarray}
These are easily integrated, but a few cases have to be considered separately.
One solution is $x_{0,t} = y_{0,t} = z_{0,t} = 0$, but this is the spherically
symmetric Stephani model that we investigated in Sec. \ref{spherlps}. So we
assume that at least one of the functions $(x_0, y_0, z_0)$ is non-constant. By
a coordinate transformation we may choose this to be $x_0$. When $x_{0,t} \neq
0$, (\ref{a.134}) is integrated with the result
\begin{equation}\label{a.137}
\frac 1 {2K} = D_1 x_0 + D_2,
\end{equation}
and then we consider

\medskip

{\bf \underline {Case 2.1: $D_1 \neq 0$}}

\medskip

In this case, (\ref{a.135}) -- (\ref{a.136}) imply
\begin{equation}\label{a.138}
y_0 = C_2 x_0, \qquad z_0 = C_3 x_0,
\end{equation}
with zero values of $C_2$ and $C_3$ allowed.

The terms free of $(x, y, z)$ in the equations $G_x = Gy = G_z = 0$ are all the
same:
\begin{eqnarray}\label{a.139}
&& \left(\frac 1 R\right),_t K,_{tt} - \left(\frac 1 R\right),_{tt} K,_t
\nonumber \\
&& - 2 K K,_t \left({x_{0,t}}^2 + {y_{0,t}}^2 + {z_{0,t}}^2\right) = 0.
\end{eqnarray}
Likewise, the terms linear in $(x, y, z)$ are the same in all 3 equations,
namely:
\begin{eqnarray}\label{a.140}
&& K \left(\frac 1 R\right),_{tt} x_{0,t} - K \left(\frac 1 R\right),_t x_{0,tt}
- 2 K,_t \left(\frac 1 R\right),_t x_{0,t} \nonumber \\
&& + 2 K^2 x_{0,t} \left({x_{0,t}}^2 + {y_{0,t}}^2 + {z_{0,t}}^2\right) = 0,
\end{eqnarray}
and the analogous equations with $x_{0,t}$ replaced by $y_{0,t}$ and $z_{0,t}$.

Using (\ref{a.137}) and (\ref{a.138}), eqs. (\ref{a.139}) and (\ref{a.140})
become identical and are integrated with the result:
\begin{equation}\label{a.141}
\frac 1 R \left(D_1 x_0 + D_2\right) + \left(1 + {C_2}^2 + {C_3}^2\right)
{x_0}^2 + C_4 x_0 = C_5.
\end{equation}
By an orthogonal coordinate transformation in the $(x, y, z)$ space one can
obtain $C_2 = C_3  = 0$. Then the model is seen to be axially symmetric, but is
more general than the plane- or hyperbolically symmetric subcases.

\medskip

{\bf \underline {Case 2.2: $D_1 = 0$}}

\medskip

Then $K = 1 / (2D_2) =$ constant, and (\ref{a.134}) -- (\ref{a.136}) and
(\ref{a.139}) are fulfilled identically. Instead of (\ref{a.140}) and its
associated equations we then have their subcases $K,_t = 0$ which imply either
$x_{0,t} = y_{0,t} = z_{0,t} = 0$ or (\ref{a.138}) plus the integral of
(\ref{a.140}), which is
\begin{equation}\label{a.142}
\frac 1 R + \frac 1 {2D_2} \left(1 + {C_2}^2 + {C_3}^2\right) {x_0}^2 + E_1 x_0
= E_2.
\end{equation}
This is the subcase $D_1 = 0$ of (\ref{a.141}).

\bigskip

{\bf Acknowledgments:} This work was partly supported by the Polish Ministry of
Higher Education grant N N202 104 838.

\end{document}